\documentclass[fleqn,usenatbib]{mnras}

\usepackage{newtxtext,newtxmath}
\usepackage[T1]{fontenc}
\usepackage{ae,aecompl}

\DeclareRobustCommand{\VAN}[3]{#2}
\let\VANthebibliography\thebibliography
\def\thebibliography{\DeclareRobustCommand{\VAN}[3]{##3}\VANthebibliography}

\usepackage{graphicx}
\usepackage{amsmath}
\usepackage{threeparttable}
\usepackage{lscape}
\usepackage[separate-uncertainty=true,multi-part-units=single]{siunitx}
\usepackage{titlesec}
\usepackage{xcolor}
\usepackage[capitalize,noabbrev]{cleveref}

\DeclareSIUnit\percent{per~cent}
\DeclareSIUnit\jansky{Jy}
\DeclareSIUnit\beam{PSF}
\DeclareSIUnit\parsec{pc}
\DeclareSIUnit\erg{erg}
\DeclareSIUnit\year{yr}
\DeclareSIUnit{\radlum}{\erg\per\second\per\hertz}
\DeclareSIUnit{\mjpb}{\milli\jansky\per\beam}

\newcommand{\diff}{\,\mathrm{d}}
\newcommand{\fracpol}{|S_V|/S_I}
\newcommand{\surveyarea}{\SI{5131}{\deg^2}}
\newcommand{\horizon}{\SI{25}{\parsec}}
\newcommand{\errorrange}[3]{{#1}^{\thinspace+{#2}}_{-{#3}}}
\newcommand{\longtermdkmdensity}{$\errorrange{9}{11}{7}\times 10^{-3}$ \SI{}{\per\deg^{-2}}}
\newcommand{\stardensity}{$1.7 \pm 0.2 \times 10^{-3}$ \SI{}{\deg^{-2}}}
\newcommand{\dkmdensity}{$\errorrange{7.1}{1.6}{1.4}\times 10^{-4}$ \SI{}{\deg^{-2}}}
\newcommand{\dfmean}{\SI{8.5}{\percent}}
\newcommand{\vaststarcount}{${\sim}200 \pm 50$}
\newcommand{\vastdkmstarcount}{${\sim}130 \pm 40$}
\newcommand{\skastardensity}{$1.2 \pm 0.3$ \SI{}{\deg^{-2}}}
\newcommand{\skadkmdensity}{$0.7 \pm 0.2$ \SI{}{\deg^{-2}}}
\newcommand{\skastarcount}{$\errorrange{41\,000}{10\,000}{9\,000}$}
\newcommand{\dkmstarprediction}{$\errorrange{47}{16}{13}$}
\newcommand{\dkmactivityfraction}{$10 \pm 3$ \si{\percent}}
\newcommand{\mean}[1]{\langle #1 \rangle}

\newcommand{\citeg}[1]{\citep[e.g.][]{#1}}

\title[VAST Pilot Stokes V]{Multi-epoch sampling of the radio star population with the Australian SKA Pathfinder}

\author[Joshua Pritchard et al.]
{
Joshua Pritchard,$^{1,2,3}$\thanks{Email: joshua.pritchard@sydney.edu.au}
Tara Murphy,$^{1,3}$\thanks{Email: tara.murphy@sydney.edu.au}
George Heald,$^{4}$
Michael S.\ Wheatland,$^{1}$
\newauthor{}
David L.\ Kaplan,$^{5}$
Emil Lenc,$^{2}$
Andrew O'Brien,$^{2,5}$
Ziteng Wang$^{1,3}$
\\
$^{1}$Sydney Institute for Astronomy, School of Physics, University of Sydney, NSW 2006, Australia\\
$^{2}$CSIRO Space and Astronomy, PO Box 76, Epping, NSW 1710, Australia\\
$^{3}$ARC Centre of Excellence for Gravitational Wave Discovery (OzGrav), Hawthorn, Victoria, Australia \\
$^{4}$CSIRO Space and Astronomy, PO Box 1130, Bentley, WA 6102, Australia \\
$^{5}$Department of Physics, University of Wisconsin-Milwaukee, P.O. Box 413, Milwaukee, WI 53201, USA \\
}

\date{Accepted XXX. Received YYY; in original form ZZZ}

\pubyear{2023}

\begin{document}
\label{firstpage}
\pagerange{\pageref{firstpage}--\pageref{lastpage}}
\maketitle

\begin{abstract}
  The population of radio-loud stars has to date been studied primarily through either targeted
  observations of a small number of highly active stars or widefield, single-epoch surveys that
  cannot easily distinguish stellar emission from background extra-Galactic sources. As a
  result it has been difficult to constrain population statistics such as the surface density
  and fraction of the population producing radio emission in a particular variable or spectral
  class. In this paper we present a sample of 36 radio stars detected in a circular
  polarisation search of the multi-epoch Variables and Slow Transients (VAST) pilot survey with
  ASKAP at \SI{887.5}{\mega\hertz}. Through repeat sampling of the VAST pilot survey footprint
  we find an upper limit to the duty cycle of M-dwarf radio bursts of \dfmean, and that
  at least \dkmactivityfraction\ of the population should produce radio bursts more luminous
  than \SI{e15}{\radlum}. We infer a lower limit on the long-term surface density of such
  bursts in a shallow \SI{1.25}{\mjpb} sensitivity survey of \longtermdkmdensity\ and an
  instantaneous radio star surface density of \stardensity\ on \SI{12}{\minute} timescales.
  Based on these rates we anticipate \vaststarcount\ new radio star detections per year over
  the full VAST survey and \skastarcount\ in next-generation all-sky surveys with the Square
  Kilometre Array.
\end{abstract}


\section{INTRODUCTION}

A wide variety of stellar systems produce magnetically driven non-thermal radio emission,
featuring a high degree of fractional circular polarisation and large brightness
temperatures. In analogy to solar radio activity, K- and M-type dwarfs produce slowly varying
or ``quiescent'' emission as well as stochastic bursts associated with magnetic reconnection
and space weather \citeg{Villadsen2019, Zic2020}. Close-in RS Canum Venaticorum (RS CVn) and
Algol binary systems produce both quiescent and bursty emission associated with the coupling of
their magnetospheres \citeg{Morris1988, White1995, Toet2021}, which feature strong magnetic
fields due to tidally induced rapid rotation of the component stars. Non-thermal emission has
also been observed from young stellar objects (YSOs), as magnetic structures connecting the
star and proto-planetary disk undergo co-rotation breakdown and reconnect \citeg{Andre1996}. B-
and A-type main-sequence stars are too hot to support a convective zone and hence lack a
conventional $\alpha-\Omega$ or $\alpha^2$ dynamo mechanism to generate strong magnetic fields;
however highly circularly polarised pulses are none the less observed from a small number of
strongly magnetic early-type chemically peculiar stars (MCPs) \citeg{Trigilio2000,
  Das2022a}. See reviews by \citet{Gudel2002} and \citet{Matthews2019} for comprehensive
overviews of the radio properties of each of these systems.

In addition to Solar-type radio activity, cool stars are known to produce highly circularly
polarised pulses of radio emission analogous to Jupiter's auroral decametric radiation. These
pulses have been observed from early M-dwarfs \citeg{Zic2019, Bastian2022} to ultracool dwarfs
\citeg{Hallinan2006, Lynch2015, Dobie2023} and brown dwarfs \citeg{Route2012, Williams2015,
  Kao2016, Kao2018, Vedantham2020b, Vedantham2023a, Rose2023} at the end of the main sequence
where stellar interiors become fully convective. Auroral pulses are widely believed to be
generated by the electron cyclotron maser instability \citep[ECMI;][]{Treumann2006}, a coherent
emission process in which gyro-phase angle bunching of electrons along with a positive gradient
in the electron velocity distribution provides free energy to drive the amplification of
circularly polarised radio emission. The ECMI operates at the local relativistic cyclotron
frequency $\omega_c = \frac{eB}{\gamma m_{e}c}$ where $e$ is the elementary charge, $m_e$ is
the electron mass, $B$ is the magnetic field strength, $\gamma$ is the Lorentz factor, and $c$
is the speed of light. Auroral radio pulses thus offer a direct measurement of the magnetic
field strength in the source region, providing insights into the origin and evolution of
magnetic fields in cool stars. This information helps to constrain the plausible dynamo
mechanisms responsible for generating strong magnetic fields \citeg{Kao2016}, map the
configuration of active coronal loops in the magnetosphere \citeg{Lynch2015}, and inform models
of the evolution and dissipation of magnetic fields by stellar winds and space weather
\citeg{Vidotto2012}.

Most of the previous studies of stellar radio activity have been limited to targeted
observations and monitoring campaigns of stars with previously identified indicators of
magnetic activity; such as strong radio activity \citeg{Villadsen2019}, flaring or variability
in optical, ultra-violet, and X-ray bands \citeg{White1989b}, or the presence of chromospheric
emission or absorption lines \citeg{Slee1987}. While these studies are ideal for modeling the
complex electrodynamic environments that drive the radio properties of individual systems, the
inherent selection bias prevents inference of statistical properties in the unobserved
population, such as the burst luminosity and rate distributions, and the fraction of the
population producing radio activity. For example \citet{White1989b} found a radio-loud fraction
of \SI{40}{\percent} in a survey of optical flare stars within \SI{10}{\parsec} at
\SI{6}{\centi\meter} and \SI{20}{\centi\meter} wavelengths, though this result may
over-estimate the radio-loud fraction by ignoring the many late-type stars without optical
flare activity that have been discovered to demonstrate similar radio burst phenomena
\citeg{Pritchard2021, Callingham2021b}. \citet{Villadsen2019} surveyed a sample of five active
M-dwarfs at \SIrange{1}{1.4}{\giga\hertz} over 13 epochs and detected 22 coherent radio bursts
inferring a burst duty cycle of \SI{25}{\percent}, though again the behaviour of these highly
active targets is difficult to extrapolate to the unobserved population.

A small number of widefield, untargeted searches for radio stars have been conducted, though
the high surface density of active galactic nuclei (AGN) typically results in a large number of
false-positive matches to foreground stars. \citet{Helfand1999} searched \SI{5000}{\deg^2} of
high Galactic latitude sky to a sensitivity of \SI{0.7}{\mjpb} in the VLA Faint Images of the
Radio Sky at Twenty-cm \citep[FIRST;][]{Becker1995} survey identifying 26 radio
stars. \citet{Kimball2009} further explored FIRST identifying 112 matches to spectrally
confirmed stars in the Sloan Digital Sky Survey \citep[SDSS;][]{Adelman-McCarthy2008}, though a
similar number of matches are estimated due to chance alignment with background radio
galaxies. \citet{Vedantham2020a} crossmatched radio sources detected by the LOw-Frequency ARray
\citep[LOFAR;][]{vanHaarlem2013} in the LOFAR Two-Metre Sky Survey
\citep[LoTSS;][]{Shimwell2017} with nearby stars in Gaia Data Release 2 \citep{Andrae2018},
discovering an M-dwarf producing coherent, circularly polarised auroral emission at metre
wavelengths.

Circular polarisation searches have been demonstrated as an excellent method for unambiguous
identification of stellar radio bursts, as the synchrotron emission produced by AGN is
typically at most \SI{2}{\percent} circularly polarised \citep{Macquart2002}, and therefore the
false-positive associations that dominate widefield searches in total intensity are negligible
in circular polarisation. The first all-sky circular polarisation transient survey was
conducted by \citet{Lenc2018} at \SI{200}{\mega\hertz} with the Murchison Widefield Array
\citep[MWA;][]{Bowman2013}, and resulted in detection of 33 previously known pulsars and two
tentative star detections. We previously reported results \citep{Pritchard2021} from an all-sky
circular polarisation search at \SI{887.5}{\mega\hertz} with the Australian Square Kilometre
Array Pathfinder \citep[ASKAP;][]{Johnston2008,Hotan2021} as part of the low-band Rapid ASKAP
Continuum Survey \citep[RACS-low;][]{McConnell2020}, with detection of 33 stars including 23
which had not been previously detected. \citet{Callingham2023} conducted a widefield circular
polarisation search at \SI{144}{\mega\hertz} as part of LoTSS with detection of 37 stars,
including a sample of 19 radio loud M-dwarfs \citep{Callingham2021b} and 14 RS CVn systems
\citep{Toet2021}.

Due to the variable nature of stellar radio bursts, population studies are fundamentally
limited in single epoch surveys, where a non-detection may either represent a radio quiet star
or simply a low burst duty cycle and thus only provide an upper limit on the burst rate of
${\sim}1/T_{\rm obs}$ where $T_{\rm obs}$ is the observing time. Telescopes such as LOFAR,
ASKAP, and the MWA have large fields of view enabling widefield single epoch surveys with a
significant overlap in sky coverage between individual pointings \citeg{Lenc2018,
  Pritchard2021, Callingham2021b, Callingham2023, Duchesne2023a}, and therefore some
variability information is provided for stars detected in these regions.  However without
dedicated multi-epoch observations these surveys are only able to acquire a small number of
repeat samples within a subset of the survey footprint, making it difficult to extrapolate
detection rates to the undetected population. In contrast multi-epoch surveys can both directly
measure the high end of the burst rate distribution from repeat detection counts, and more
strongly constrain the low end through statistical analysis of non-detection rates. These
parameters are important inputs to forecasts of the surface density of stellar radio bursts in
current and next-generation radio transient surveys.

In this paper we present the results of a multi-epoch circular polarisation search for radio
stars across RACS-low and the low-band of the VAST Pilot Survey
\citep[VASTP-low;][]{Murphy2021}. In \cref{sec:data} we summarise the RACS-low and VASTP-low
observations and data processing; in \cref{sec:search} we describe our circular polarisation
search procedure and candidate selection; in \cref{sec:radio-properties} we present the radio
luminosity and polarisation properties of our sample of detected stars; and in
\cref{sec:popstats} we use the subset of stellar detections within the repeat-sampled VASTP-low
footprint to derive statistical properties of the radio loud M-dwarf population.

\section{OBSERVATIONS AND DATA REDUCTION}\label{sec:data}

\subsection{Standard Processing}\label{sec:askapsoft}

\begin{table}
  \centering
  \caption{Summary of RACS-low and VASTP-low observing parameters.}\label{tab:observations}
    \begin{tabular}[h!]{lrr}
      \hline
      Property                         & Value & Survey \\
      \hline
      Central Frequency                & \SI{887.5}{\mega\hertz} & \\
      Bandwidth                        & \SI{288}{\mega\hertz} & \\
      Integration Time                 & \SI{12}{\minute} & VASTP-low \\
                                       & \SI{15}{\minute} & RACS-low \\
      Median RMS Noise                 & \SI{240}{\micro\jansky\per\beam} & VASTP-low \\
                                       & \SI{250}{\micro\jansky\per\beam} & RACS-low \\
      Polarisation Leakage             & \SI{0.6}{\percent} (V positive) & VASTP-low \\
                                       & \SI{0.7}{\percent} (V negative) & VASTP-low \\
                                       & \SI{0.5}{\percent} (V positive) & RACS-low \\
                                       & \SI{0.6}{\percent} (V negative) & RACS-low \\
      Astrometric Accuracy             & $\Delta\alpha\cos\delta$ = \SI{-0.19\pm 0.53}{\arcsec} & VASTP-low \\
                                       & $\Delta\delta$ = \SI{0.07\pm 0.48}{\arcsec} & VASTP-low \\
                                       & $\Delta\alpha\cos\delta$ = \SI{-0.60\pm 0.50}{\arcsec} & RACS-low \\
                                       & $\Delta\delta$ = \SI{+0.10\pm 0.50}{\arcsec} & RACS-low \\
      PSF Central Lobe (FWHM)          & $B_{\rm min}$ = \SI{11.3\pm 0.6}{\arcsec} & VASTP-low \\
                                       & $B_{\rm max}$ = \SI{13.7\pm 4.2}{\arcsec} & VASTP-low \\
                                       & $B_{\rm min}$ = \SI{11.8\pm 0.9}{\arcsec} & RACS-low \\
                                       & $B_{\rm max}$ = \SI{18.0\pm 4.3}{\arcsec} & RACS-low \\
      \hline
    \end{tabular}
\end{table}

Full details of RACS-low and VASTP-low observations and standard processing are provided in
\citet{McConnell2020} and \citet{Murphy2021} respectively, but we summarise them
here. Observations from both RACS-low and VASTP-low use a $6\times6$-beam square footprint with
a \SI{66}{\deg^2} field of view at a central frequency of \SI{887.5}{\mega\hertz} with 288
channels each \SI{1}{\mega\hertz} wide. RACS-low and VASTP-low observations were acquired with
\SI{15}{\minute} and \SI{12}{\minute} integrations reaching a median RMS noise $\sigma_I$ of
\SI{250}{\micro\jansky\per\beam} and \SI{240}{\micro\jansky\per\beam} respectively. The
observing parameters are summarised in \cref{tab:observations}, and the temporal and sky
coverage of each epoch is listed in \cref{tab:epochs}.

Each ASKAP antenna is equipped with a Phased Array Feed \citep[PAF;][]{Hotan2014,
  McConnell2016} which allows the formation of 36 dual linear polarisation beams on the sky.
All four cross-correlations are recorded allowing calibration of the frequency-dependent XY
phases and full reconstruction of Stokes $I$, $Q$, $U$, and $V$ images. The antenna roll-axis
is adjusted throughout each observation to maintain orientation of the linear feeds with
respect to the celestial coordinate frame so that no correction for parallactic angle is
required, and keeping the beam footprint fixed on the sky. ASKAP data are calibrated and imaged
with the {\sc ASKAPsoft} package \citep{Cornwell2011, Guzman2019} which generates full Stokes
image products and uses the {\sc selavy} \citep{Whiting2012b} source finder package to extract
a catalogue of 2D Gaussian source components from the Stokes $I$ images.  We determined the
signal-to-noise dependent astrometric uncertainty of extracted {\sc selavy} components
following \citet{Condon1997} as

\begin{equation}\label{eq:astrometric-uncertainty}
  \sigma_{\theta} = \frac{2\theta_m}{\frac{S_I}{\sigma_I}\sqrt{8\ln{2}}},
\end{equation}

where $\theta_m$ is the component major axis and $\sigma_I$ is the local RMS noise. We added
this uncertainty in quadrature to the astrometric accuracies listed in \cref{tab:observations}
to determine the positional uncertainty of each radio source, with uncertainties of $2\farcs3$
and $3\farcs1$ for a $5\sigma_I$ detection in RACS-low and VASTP-low respectively.

\subsection{Stokes V Processing}\label{sec:stokesv-processing}

We ran {\sc selavy} on the Stokes $V$ images with standard {\sc ASKAPsoft} settings, with two
extractions to collect both positive and negative sources. Throughout this paper we adopt the
IAU/IEEE convention for Stokes $V$, in which positive and negative Stokes $V$ correspond to
right and left handed circular polarisation respectively \citep{Robishaw2018}. We checked that
our fluxes are consistent with this convention by comparison to a selection of 25 pulsars
reported in \citet{Han1998} and \citet{Johnston2018} which are persistently detected with the
same circular polarisation handedness at ${\sim}\SI{1}{\giga\hertz}$. All 25 pulsars are in
agreement with our measured polarisation handedness after accounting for the opposite
convention used in pulsar astronomy. We also checked the consistency of Stokes $V$ sign across
the duration of the pilot survey through the star HR~1099, an RS CVn binary system that is
detected in every epoch and maintains a negative Stokes $V$ sign in all detections. This is
consistent with the expected polarisation of gyrosynchrotron emission from this system which is
left handed below ${\sim}\SI{1.5}{\giga\hertz}$ \citep{White1995}.

We characterised the degree of polarisation leakage by crossmatching selavy components in
Stokes $I$ with their counterpart in Stokes $V$, requiring each match to have no neighbouring
components within \SI{30}{\arcsec}, Stokes $I$ flux density $S_I$ of \SI{>100}{\mjpb}, and
Stokes $V$ flux density $S_V$ greater than $5\sigma_V$ where $\sigma_V$ is the local Stokes $V$
RMS noise. We removed sources associated with stars and pulsars so that the sampled points are
associated with AGN with no intrinsic circular polarisation, and are thus a measurement of the
degree of polarisation leakage from Stokes $I$ into Stokes~$V$.

\begin{table}
  \centering
  \caption{Observing dates and sky coverage of VASTP-low epochs.}\label{tab:epochs}
  \begin{threeparttable}
    \begin{tabular}[h!]{cllcc}
      \multicolumn{1}{c}{Epoch}
      & \multicolumn{1}{c}{Start Date}
      & \multicolumn{1}{c}{End Date}
      & \multicolumn{1}{c}{Area ($\deg^2$)}
      & \multicolumn{1}{c}{Fields} \\
      \hline
      \hspace{4pt}0$^a$ & 2019 Apr 25 & 2019 May 03 & 5\,131 & $113$     \\
      1                 & 2019 Aug 27 & 2019 Aug 28 & 5\,131 & $113$     \\
      2                 & 2019 Oct 28 & 2019 Oct 31 & 4\,905 & $108$     \\
      \hspace{4pt}3x    & 2019 Oct 29 & 2019 Oct 29 & 2\,168 & $~~43$    \\
      \hspace{4pt}4x    & 2019 Dec 19 & 2019 Dec 19 & 1\,672 & $~~34$    \\
      \hspace{4pt}5x    & 2020 Jan 10 & 2020 Jan 11 & 3\,818 & $~~81$    \\
      \hspace{4pt}6x    & 2020 Jan 11 & 2020 Jan 12 & 2\,400 & $~~49$    \\
      \hspace{4pt}7x    & 2020 Jan 16 & 2020 Jan 16 & 1\,666 & $~~33$    \\
      8                 & 2020 Jan 11 & 2020 Feb 01 & 5\,097 & $112$     \\
      9                 & 2020 Jan 12 & 2020 Feb 02 & 5\,097 & $112$     \\
      10x               & 2020 Jan 17 & 2020 Feb 01 & $~~$803 & $~~13$  \\
      11x               & 2020 Jan 18 & 2020 Feb 02 & $~~$695 & $~~11$  \\
      12\hspace{4pt}    & 2020 Jan 19 & 2020 Jun 21 & 5\,100 & $112$     \\
      13\hspace{4pt}    & 2020 Aug 28 & 2020 Aug 30 & 5\,028 & $104$     \\
      17\hspace{4pt}    & 2021 Jul 21 & 2021 Jul 24 & 5\,131 & $~~114^b$ \\
      19\hspace{4pt}    & 2021 Aug 20 & 2021 Aug 24 & 5\,131 & $~~115^b$ \\
      \hline
    \end{tabular}
    \begin{tablenotes}
    \item[a] Listed values for epoch 0 are for the 113 RACS-low fields within the VASTP-low
      footprint.
    \item[b] Epochs 17 and 19 include one and two repeat observations respectively, but have
      the same total sky coverage as epoch 1.
    \end{tablenotes}
  \end{threeparttable}
\end{table}

\begin{figure}
  \centering
  \includegraphics[width=0.5\textwidth]{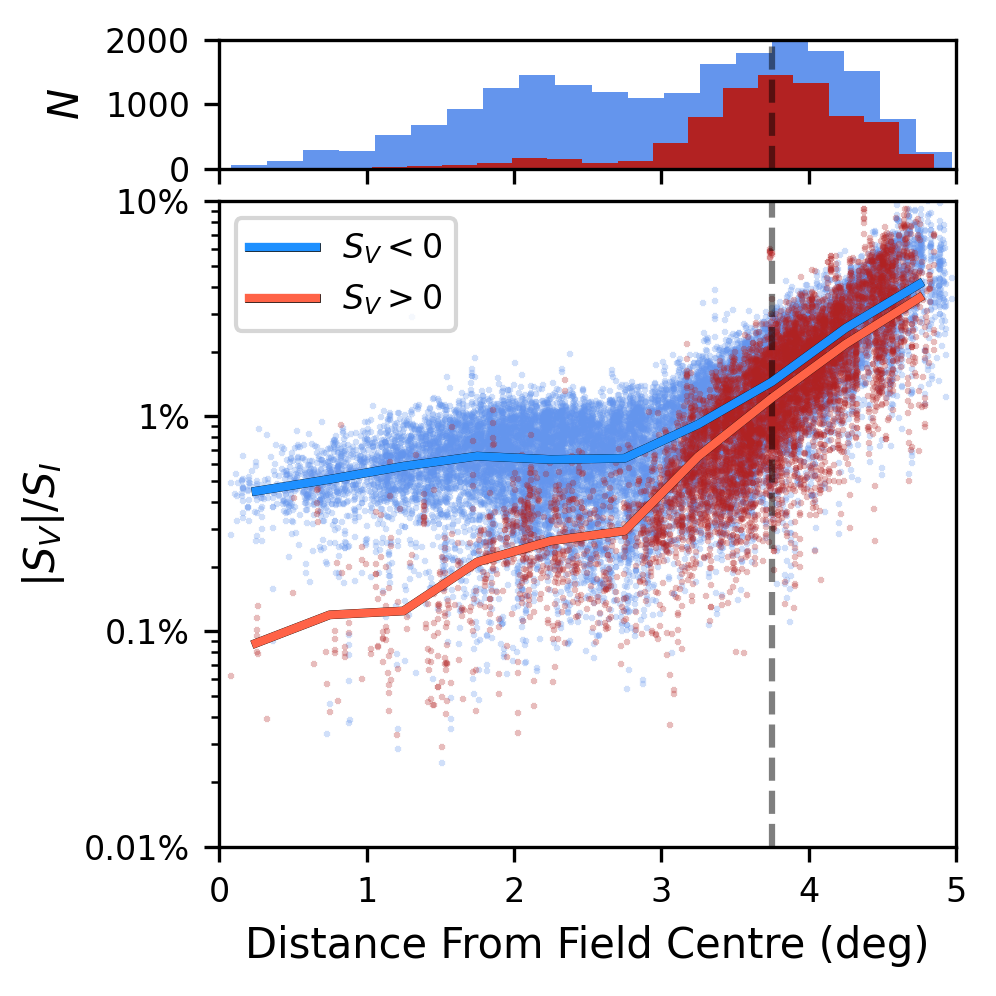} 
  \caption{
    \small
    Leakage from Stokes $I$ into $V$ ($\fracpol$) as a function of angular
    distance from field centre. The points represent Stokes $I$-$V$ matches between isolated
    point sources with $S_I > \SI{100}{\mjpb}$ with red and blue corresponding to matches with
    positive and negative Stokes $V$ flux respectively. The histograms show the distribution of
    positive and negative matches with angular distance with a bias towards negative leakage
    towards the field centre. The black dashed line indicates a field centre distance of
    \SI{3.75}{\degree}, within which the positive and negative leakage has median values of
    \SI{0.6}{\percent} and \SI{-0.7}{\percent} respectively.
  }\label{fig:leakage-fcd}
\end{figure}

We made 28\,149 leakage measurements from all images within the survey with 7\,852 positive and
20\,297 negative Stokes $V$ measurements respectively. In \cref{fig:leakage-fcd} we show the
fractional circular polarisation of each leakage measurement as a function of angular distance
to field centre along with the median leakage in 10 field centre distance bins. The larger
count of negative leakage measurements is visible as a bias towards negative Stokes $V$ leakage
near to field centre at a level of $\sim$\SI{0.5}{\percent} with very few positive leakage
measurements in this range.  At distances of greater than $\sim$\SI{3}{\degree} from field
centre both positive and negative leakage begin to increase to a level of
$\sim$\SI{1}{\percent} at $\sim$\SI{4}{\degree} towards the field edges, and
$\sim$\SI{3}{\percent} at $\sim$\SI{5}{\degree} in the corners.

\begin{figure*}
  \centering
  \includegraphics[width=\textwidth]{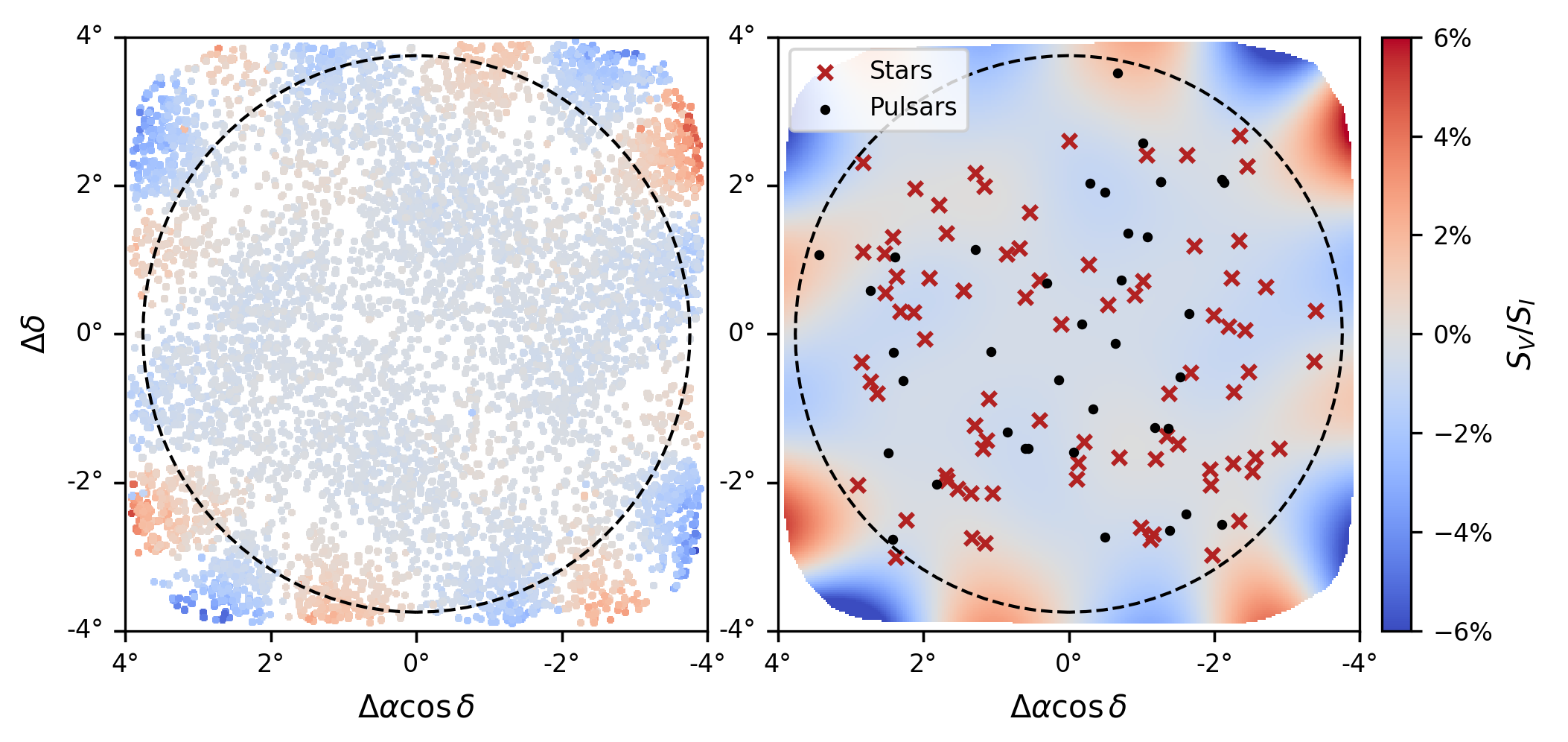} 
  \caption{
    \small
    Map of leakage from Stokes $I$ into Stokes $V$ across the ASKAP field
    of view in VASTP-low. The left panel shows position and leakage of Stokes $I$-$V$ matches
    in the field-centred frame. The right panel shows a Gaussian Process regression model fit
    to these measurements, with the positions of all star and pulsar detections overlaid. The
    black dashed circle indicates an angular distance from the field centre of
    \SI{3.75}{\degree}.
  }\label{fig:leakage-map}
\end{figure*}

We transformed the position of each sample into the image-centred frame to visualise the
leakage pattern across the ASKAP field of view. In \cref{fig:leakage-map} we show these samples
alongside a Gaussian Process regression fit to the data with the position of all star and
pulsar detections overlaid. Within the central \SI{3.75}{\degree} from field centre the leakage
has median positive and negative values of \SI{0.6}{\percent} and \SI{-0.7}{\percent}. In this
region the leakage pattern towards the edge of individual beams overlaps and suppresses the
mosaiced leakage, with a small bias towards negative Stokes~$V$ caused by imperfect overlap due
to uncertainty in the primary beam response pattern. The tile edges have no overlapping beams
and a residual leakage pattern is visible alternating between positive and negative
Stokes~$V$. We note that more recent ASKAP observations such as the mid-band epoch of RACS at
\SI{1367.5}{\mega\hertz} \citep[RACS-mid;][]{Duchesne2023a} implement widefield leakage
corrections that remove both the bias and residual leakage pattern.

\section{CANDIDATE SEARCH}\label{sec:search}

\subsection{Candidate Selection}\label{sec:selection}

We generated associations between the extracted Stokes $I$ and $V$ components with a
many-to-many crossmatch using a match radius of \SI{6}{\arcsec}. Many-to-many association
allows for one Stokes $I$ component to match to multiple Stokes $V$ components within the match
radius (and vice versa), ensuring that all possible $I$-$V$ associations within the match
radius are considered at the cost of an increased number of false-positive associations. This
choice is made to avoid missing edge cases in which a near neighbour in Stokes $I$ matches on
to the Stokes $V$ component and produces a lower $\fracpol$ than the correct Stokes $I$
component.

We then crossmatched the $I$-$V$ associations between each epoch to track individual candidates
over the duration of the survey. Beginning with an initial epoch $A$ each $I$-$V$ association
is assigned a unique association ID. We then generate an initial many-to-many crossmatch of
$I$-$V$ associations between epoch $A$ and the subsequent epoch $B$ using the Stokes $I$
coordinates with a \SI{6}{\arcsec} match radius.

Many-to-many association across $N_e$ epochs results in
$\mathcal{O}\left((n_In_V)^{N_e-1}\right)$ total candidates, with $n_I$ and $n_V$ selavy
components within the match radius in Stokes $I$ and $V$ respectively. We de-duplicate these
candidates by retaining only the $A$-$B$ associations whose match distance is the minimum of
all duplicate candidates such that there is no other copy of an epoch $A$ or $B$ $I$-$V$
association with a closer match, reducing the number of candidates to
$\mathcal{O}\left(\max(n_I,n_V)\right)$. This step greatly reduces the number of artefacts
produced around bright sources, where $I$-$V$ associations between sidelobes shift position
between epochs.

We create a running list of candidates for the epoch $A$-$B$ associations including epoch $A$
$I$-$V$ associations with no match in epoch $B$, and also add all epoch $B$ $I$-$V$
associations with no match in epoch $A$, assigning a new association ID. Finally we
re-calculate the average Stokes $I$ coordinates of each candidate and use these to perform the
next round of epoch-association. Using this procedure we performed association between the 15
epochs of VASTP-low, resulting in a list of $9\,818$ candidates. We then filtered this list to
candidates both greater than $5\sigma$ in both Stokes $I$ and $V$ where $\sigma$ is the local
RMS noise, and with at least one $I$-$V$ association with fractional circular polarisation
$\fracpol > \SI{6}{percent}$, corresponding to $10$ times the median polarisation leakage of
Stokes $I$ into $V$ and three times the typical leakage of $\sim$\SI{2}{\percent} in image
corners. Our final list contains 1\,188 candidates.

\subsection{Artefact Rejection}\label{sec:artefact-rejection}

\begin{figure*}
  \centering
  \includegraphics[width=\textwidth]{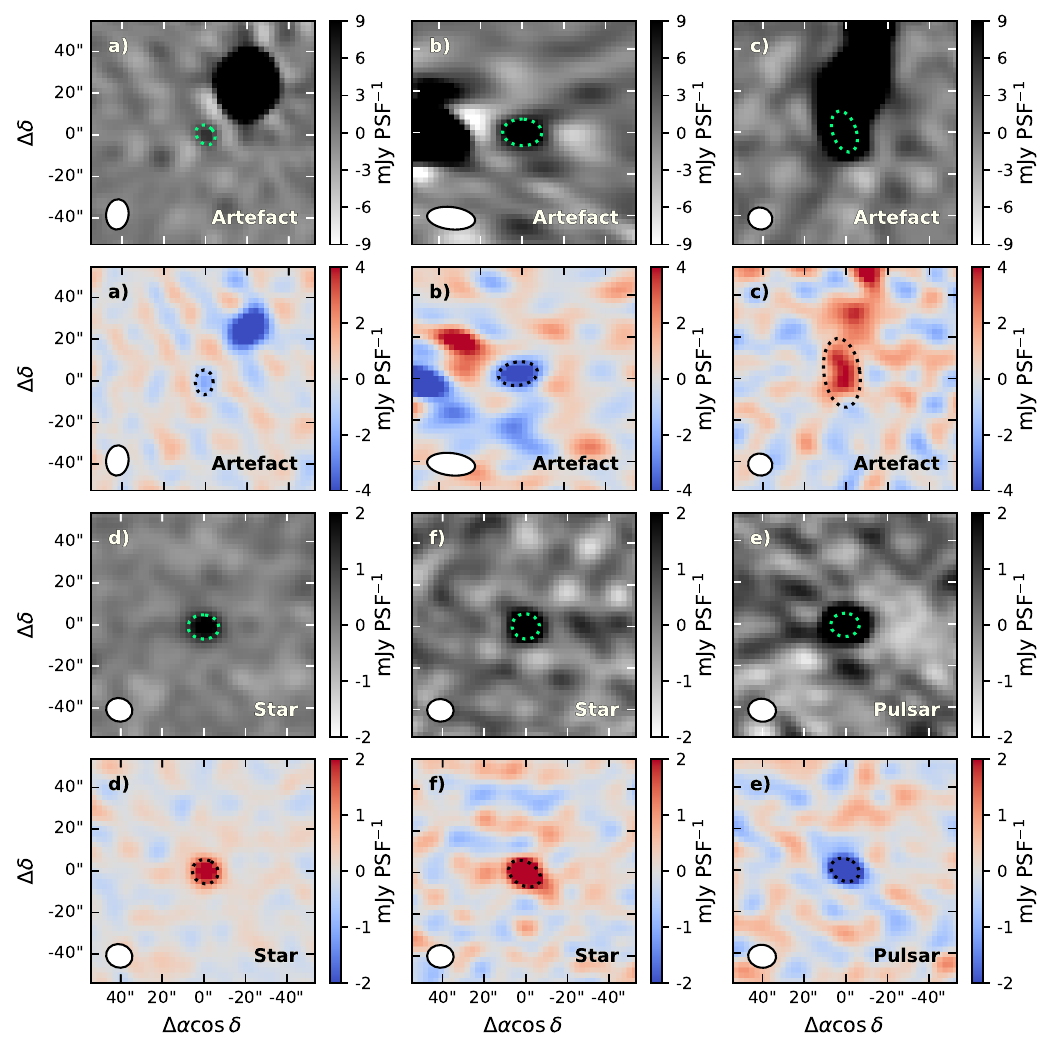}
  \caption{\small
    Stokes $I$ (top panel) and $V$ (bottom panel) cutouts of six candidates
    inspected during classification. The top pair of rows shows candidates rejected due to
    $I$-$V$ association between either the sidelobes of a bright source (panels a) and b)) or
    leakage from a multi-component extended source (panel c)). The bottom pair of rows show two
    identified stars (panels d) and e)) and a pulsar (panel f)) for comparison. The restoring
    beam is shown as the ellipse in the lower left of each panel, and green and black dotted
    ellipses trace the shape of the Stokes $I$ and $V$ {\sc selavy} components in each
    association respectively.
  }\label{fig:cutouts}
\end{figure*}

Through visual inspection of the Stokes $I$ and $V$ images we rejected $I$-$V$ associations
with artificially high $\fracpol$ caused by imaging artefacts and spurious noise. In
\cref{fig:cutouts} we show a selection of example image cutouts of rejected artefacts as well
as legitimately polarised sources. These artefacts are typically caused by one of the following
situations:

\begin{itemize}
  \item a match between a low flux component of a multi-component source in Stokes $I$
    to leakage of a brighter component in Stokes $V$,
  \item a match between components of an extended source in Stokes $I$ and leakage in
    Stokes $V$, indicated by the presence of multiple nearby components and mismatched
    component shape parameters,
  \item a match to a spurious noise peak in Stokes $V$, indicated by a combination of low
    signal to noise ratio of $< 6\sigma_V$ along with an offset between Stokes $I$ and $V$
    position of $>4 \arcsec$ and source morphology resembling noise rather than a point source,
  \item a match between the sidelobes and leakage of a poorly deconvolved bright source.
\end{itemize}

In total we removed 837 candidates based on these criteria, with the majority caused by matches
between the sidelobes of bright sources in Stokes $I$ and their leakage in Stokes $V$ as shown
in panels (a) and (b) of \cref{fig:cutouts}, or association between leakage from the
sub-components of an extended, multi-component source as shown in panel (c). We further removed
296 candidates with $\fracpol > 0.06$ caused by polarisation leakage towards the image corners,
where leakage is less well suppressed than the central region of the image.

\subsection{Classification}\label{sec:vetting}

We classified each of the remaining 54 candidates following the same procedure detailed by
\citet{Pritchard2021}; using radio and multi-wavelength image cutouts, queries to the SIMBAD
and NED databases, and radio lightcurves and spectral energy distributions derived from
archival radio data. We identified 85 total detections of 11 known pulsars, including the Large
Magellanic Cloud pulsar PSR~J0523$-$7125 discovered by \citet{WangY2022}, and seven highly
circularly polarised transients with no multi-wavelength counterpart, including the Galactic
Centre transient J1736$-$3216 discovered by \citet{WangZ2021}.

We queried SIMBAD and the Gaia Data Release 3 \citep[Gaia DR3;][]{Gaia2022} catalogue for stars
within a $3\farcm5$ radius of candidate positions. From this sample we removed sources with
parallax over error $\pi / \sigma_\pi < 10$ and distance $d < \SI{500}{\parsec}$ in order to
reject false positive alignments to distant stars in high source density regions such as the
Galactic plane. We then applied positional corrections to the radio epoch using Gaia proper
motion parameters, identifying 36 matches to stars within the SNR dependent astrometric
uncertainty determined by \cref{eq:astrometric-uncertainty}.

Combined with the stars detected in our circular polarisation search of RACS-low we have
detected a total of 76 radio stars. We quantified the rate of false-positive association
between foreground stars and extra-galactic radio sources by offsetting the positions of all 76
stars by \SIrange{5}{20}{\arcmin} in random directions and crossmatching against SIMBAD and
Gaia DR3 with the same selection criteria as above. Ten runs of this procedure resulted in zero
matches, suggesting a false-positive rate of less than \SI{0.2}{\percent} and an extremely low
probability that any of our detections are due to chance alignment.

\begin{table}
  \centering
  \caption{Count summary of radio stars in RACS-low and VASTP-low. Columns indicate the
    abbreviated class name, number count of unique stars, total individual detections
    $N_d[I]$ and $N_d[V]$ in Stokes $I$ and $V$ respectively, and total observations $N_o$ for
    each variable class of radio star.}\label{tab:starcounts}
  \resizebox{\columnwidth}{!}{
    \begin{tabular}{lrrrrrr}
      \hline
      Variable Class               &       & Stars & $N_d [I]$ & $N_d [V]$ & $N_o$ \\
      \hline
      M-dwarfs                     & (dM)  & 45    & 96       & 65        & 417   \\
      RS CVn / Algol binaries      & (IB)  & 9     & 52       & 23        & 74    \\
      Magnetic chemically peculiar & (MCP) & 7     & 17       & 7         & 17    \\
      Young stellar objects        & (YSO) & 5     & 34       & 7         & 48    \\
      Hot spectroscopic binaries   & (HSB) & 4     & 23       & 4         & 32    \\
      K-dwarfs                     & (dK)  & 4     & 5        & 5         & 16    \\
      White dwarfs                 & (WD)  & 2     & 2        & 2         & 14    \\
      \hline
    \end{tabular}
  }
\end{table}

\section{RADIO PROPERTIES}\label{sec:radio-properties}

In our combined circular polarisation searches of RACS-low and VASTP-low we have detected 76
radio stars a total of 229 times in Stokes~$I$ and 113 times in Stokes~$V$.  The majority of
detected stars are M-dwarfs (dM) or interacting RS CVn / Algol binary systems (IB), with a
small number of K-dwarfs (dK), YSOs, and MCP stars, and two white dwarfs (WD). We summarise the
detection counts of each variable class detected in RACS-low and VASTP-low in
\cref{tab:starcounts}, and in \cref{tab:startable} we list the radio properties of each star.

We detect a higher proportion of dM to IB systems in comparison to radio stars detected in
FIRST by \citet{Helfand1999} and LoTSS by \citet{Callingham2023}, whose samples are each
approximately \SI{50}{\percent} comprised of IB stars. This can be explained by the differences
in survey strategy along with the relative timescales of dM and IB radio emission, where dM
stars generally feature faint quiescent emission and radio bursts lasting seconds to hours
\citeg{Villadsen2019}, while IB systems typically have brighter, long-duration quiescent
emission \citeg{Chiuderi-Drago1990}.  FIRST observations are composed of \SI{165}{\second}
snapshots co-added to an effective duration of $\sim$\SI{35}{\minute} while the VASTP-low
survey footprint is sampled a median of 14 times in \SI{12}{\minute} snapshots, increasing the
likelihood of detecting M-dwarf bursts. Moreover, FIRST observations covered a narrow bandwidth
of \SI{42}{\mega\hertz} which may have limited the capability to detect coherent M-dwarf bursts
with sharp spectral cutoffs, while the detectability of broadband gyrosynchrotron emission
associated with IB systems is less affected by bandwidth. LoTSS images are constructed with an
integration time of \SI{8}{\hour}, and the M-dwarfs reported by \citet{Callingham2023} all
feature long-duration emission spanning the full observation \citep{Callingham2021}.  Short
time-scale bursts are likely to be missed in long integration images due to dilution of the
burst flux with lengthy periods in which burst activity is low or absent. Alternatively, the
difference in dM and IB detection counts may be explained by a change in the emission
characteristics of each source class between the LoTSS frequency of \SI{144}{\mega\hertz} and
VASTP-low frequency of \SI{887.5}{\mega\hertz}, though further analysis of the radio stars
detected in each band is required to disentangle survey strategy biases from intrinsic changes
in the emission physics.

\begin{landscape}
  \begin{table}
    \centering
\caption{
  Table of radio stars detected in the VASTP-low and RACS-low surveys. Listed columns are
  stellar name, radio variable and spectral classes, radio coordinates, proper motion
  parameters $\mu_\alpha\cos\delta$ and $\mu_\delta$, parallax $\pi$, upper limit to emission
  region length scale $R_e$, \SI{887.5}{\mega\hertz} Stokes $I$ peak flux density $S_I$,
  fractional circular polarisation $\fracpol$, lower limit to brightness temperature
  $\log_{10}T_B$, radio luminosity $\log_{10}L_\nu$, and the number of Stokes $I$ and $V$
  detections $N_d[I]$ and $N_d[V]$, and total number of observations $N_o$ of each star over
  the survey. $S_I$, $\fracpol$, $T_B$, and $L_\nu$ are reported as ranges over the detections
  of each star. A machine-readable version of this table is available online.
  }\label{tab:startable} \resizebox{\columnwidth}{!}{
\begin{threeparttable}
\begin{tabular}{lclrrrrccccccccc}
\hline
 \multicolumn{1}{c}{Name}     & \multicolumn{1}{c}{Variable Type}   & \multicolumn{1}{c}{Spectral Class}   & \multicolumn{1}{c}{RA}   & \multicolumn{1}{c}{Dec}   & \multicolumn{1}{c}{$\mu_\alpha\cos\delta$}   & \multicolumn{1}{c}{$\mu_\delta$}   & \multicolumn{1}{c}{$\pi$}   & \multicolumn{1}{c}{$R_e$\tnote{a}}   & \multicolumn{1}{c}{$S_I$}   & \multicolumn{1}{c}{$\fracpol$}   & \multicolumn{1}{c}{$\log_{10}T_B$}   & \multicolumn{1}{c}{$\log_{10}L_\nu$}   &   \multicolumn{1}{c}{$N_d[I]$} &   \multicolumn{1}{c}{$N_d[V]$} &   \multicolumn{1}{c}{$N_o$} \\
&&&&&\si{mas\per\year}&\si{mas\per\year}&\si{mas}&$R_\odot$&\si{\mjpb}&&\si{\kelvin}&\si{\radlum}&&&\\
\hline
 G~131-26                     & dM                                  & M5V                                  & 00:08:53.88              & $+$20:50:19.36            & $~-48.64$                                    & $-260.19$                          & $~~~55.26 \pm 0.76$         & $~~~1.0$                             & $2.78$                      & $0.57$                           & $10.3$                               & $15.0$                                 &                              1 &                              1 &                           1 \\
 1RXS~J001650.6-071013        & dM                                  & M0                                   & 00:16:50.11              & $-$07:10:15.19            & $~~~~39.84$                                  & $~~~~~9.74$                        & $~~~14.01 \pm 0.06$         & $~~~2.3$                             & $0.85$--$3.07$              & $0.19$--$1.00$                   & $10.3$--$10.9$                       & $15.7$--$16.3$                         &                              8 &                              4 &                          11 \\
 CF~Tuc                       & IB                                  & G2/5V+F0                             & 00:53:09.12              & $-$74:39:05.75            & $~~~243.13$                                  & $~~~~21.62$                        & $~~~11.37 \pm 0.03$         & $~~11.3$                             & $1.31$--$10.98$             & $0.06$--$0.62$                   & $9.3$--$10.2$                        & $16.1$--$17.0$                         &                             10 &                              1 &                          10 \\
 CS~Cet                       & dK                                  & K0IV(e)                              & 01:06:48.93              & $-$22:51:23.22            & $~-38.15$                                    & $~-81.19$                          & $~~~10.50 \pm 0.06$         & $~~12.3$                             & $6.32$--$7.30$              & $0.37$--$0.39$                   & $10.0$--$10.1$                       & $16.8$--$16.9$                         &                              2 &                              2 &                           2 \\
 BI~Cet                       & IB                                  & G5V                                  & 01:22:50.09              & $+$00:42:39.60            & $-114.59$                                    & $-238.95$                          & $~~~15.98 \pm 0.05$         & $~~~4.8$                             & $1.20$--$3.45$              & $0.21$--$0.99$                   & $9.7$--$10.2$                        & $15.8$--$16.2$                         &                              7 &                              2 &                           9 \\
 RX~J0143.7-0602              & dM                                  & M3.5                                 & 01:43:45.14              & $-$06:02:40.27            & $~~~~52.26$                                  & $~-27.07$                          & $~~~46.85 \pm 0.07$         & $~~~1.2$                             & $2.89$                      & $0.56$                           & $10.4$                               & $15.2$                                 &                              1 &                              1 &                          13 \\
 WISEA~J014358.01-014930.3    & dM                                  &                                      & 01:43:58.02              & $-$01:49:28.73            & $~~~~60.35$                                  & $~~~~~3.77$                        & $~~~24.80 \pm 0.02$         & $~~~1.0$                             & $1.44$                      & $0.84$                           & $10.7$                               & $15.4$                                 &                              1 &                              1 &                          12 \\
 SDSS~J020648.78-061416.3     & IB                                  &                                      & 02:06:48.76              & $-$06:14:16.30            & $~~~~~6.13$                                  & $~-24.34$                          & $~~~~2.25 \pm 0.04$         & $~~~4.8$                             & $1.64$                      & $0.96$                           & $11.6$                               & $17.6$                                 &                              1 &                              1 &                          21 \\
 UPM~J0250-0559               & dM                                  &                                      & 02:50:40.29              & $-$05:59:50.89            & $~~~~67.31$                                  & $~~~~10.17$                        & $~~~33.20 \pm 0.96$         & $~~~1.2$                             & $2.25$                      & $0.82$                           & $10.6$                               & $15.4$                                 &                              1 &                              1 &                          24 \\
 LP~~771-50                   & dM                                  & M5e                                  & 02:56:27.19              & $-$16:27:38.53            & $~~~212.29$                                  & $~-55.25$                          & $~~~32.56 \pm 0.41$         & $~~~0.6$                             & $1.65$                      & $1.00$                           & $11.1$                               & $15.3$                                 &                              1 &                              1 &                           1 \\
 CD-44~~1173                  & dM                                  & K6Ve                                 & 03:31:55.73              & $-$43:59:14.91            & $~~~~88.47$                                  & $~~-3.50$                          & $~~~22.17 \pm 0.03$         & $~~~2.5$                             & $3.87$                      & $0.88$                           & $10.5$                               & $16.0$                                 &                              1 &                              1 &                          19 \\
 HD~~22468                    & IB                                  & K2:Vnk                               & 03:36:47.19              & $+$00:35:13.52            & $~-32.89$                                    & $-161.77$                          & $~~~33.75 \pm 0.09$         & $~~~9.1$                             & $6.92$--$78.17$             & $0.03$--$0.54$                   & $9.3$--$10.3$                        & $15.9$--$16.9$                         &                             13 &                             13 &                          13 \\
 RX~J0348.9+0110              & dK                                  & K3V:/E                               & 03:48:58.83              & $+$01:10:54.83            & $~~~~29.73$                                  & $~-22.72$                          & $~~~~9.11 \pm 0.33$         & $~~~0.9$                             & $1.42$                      & $0.93$                           & $11.8$                               & $16.3$                                 &                              1 &                              1 &                          12 \\
 HD~~24681                    & YSO                                 & G5V                                  & 03:55:20.61              & $-$01:43:46.58            & $~~~~43.33$                                  & $~-91.49$                          & $~~~17.82 \pm 0.04$         & $~~~2.6$                             & $1.88$                      & $0.54$                           & $10.3$                               & $15.9$                                 &                              1 &                              1 &                          13 \\
 HD~~24916B                   & dM                                  & M2.5V                                & 03:57:28.66              & $-$01:09:27.07            & $-209.41$                                    & $-139.73$                          & $~~~65.49 \pm 0.04$         & $~~~1.3$                             & $1.02$                      & $0.91$                           & $9.5$                                & $14.5$                                 &                              1 &                              1 &                          13 \\
 UPM~J0409-4435               & dM                                  &                                      & 04:09:32.15              & $-$44:35:38.19            & $~-16.86$                                    & $~~~115.59$                        & $~~~68.07 \pm 0.04$         & $~~~0.7$                             & $1.27$--$2.42$              & $0.73$--$1.00$                   & $10.2$--$10.4$                       & $14.5$--$14.8$                         &                              3 &                              1 &                          23 \\
 RX~J0419.2-7120              & dM                                  &                                      & 04:19:13.35              & $-$71:21:12.05            & $~-52.92$                                    & $~-46.71$                          & $~~~18.77 \pm 0.13$         & $~~~1.1$                             & $1.45$                      & $0.58$                           & $10.9$                               & $15.7$                                 &                              1 &                              1 &                           1 \\
 HD~283750                    & dK                                  & K2.5Ve                               & 04:36:48.60              & $+$27:07:55.01            & $~~~232.87$                                  & $-148.14$                          & $~~~57.10 \pm 0.06$         & $~~~2.0$                             & $3.94$                      & $0.76$                           & $9.9$                                & $15.2$                                 &                              1 &                              1 &                           1 \\
 CD-56~~1032                  & dM                                  & M3Ve+M4Ve                            & 04:53:31.02              & $-$55:51:35.10            & $~~~126.12$                                  & $~~~~76.86$                        & $~~~90.12 \pm 0.03$         & $~~~1.3$                             & $0.96$--$12.38$             & $0.16$--$0.98$                   & $9.3$--$10.4$                        & $14.1$--$15.3$                         &                             14 &                              2 &                          23 \\
 HD~~32595                    & HSB                                 & B8                                   & 05:04:49.06              & $+$13:18:32.86            & $~~~~~1.83$                                  & $~~-5.18$                          & $~~~~3.40 \pm 0.28$         & $~~~6.5$                             & $3.36$                      & $0.62$                           & $11.2$                               & $17.5$                                 &                              1 &                              1 &                           1 \\
 V1154~Tau                    & IB                                  & B5                                   & 05:05:37.70              & $+$23:03:41.59            & $~~~~~0.27$                                  & $~~-6.89$                          & $~~~~3.59 \pm 0.46$         & $~~~8.9$                             & $2.07$                      & $0.64$                           & $10.7$                               & $17.3$                                 &                              1 &                              1 &                           1 \\
 AB~Dor                       & YSO                                 & K0V                                  & 05:28:44.98              & $-$65:26:51.35            & $~~~~29.15$                                  & $~~~164.42$                        & $~~~65.32 \pm 0.14$         & $~~~4.5$                             & $1.83$--$7.73$              & $0.16$--$1.00$                   & $8.7$--$9.4$                         & $14.7$--$15.3$                         &                             29 &                              2 &                          31 \\
 HD~~36150                    & MCP                                 & A5III/IV                             & 05:29:41.78              & $-$00:48:07.66            & $~~~~71.34$                                  & $~-47.04$                          & $~~~~8.93 \pm 0.18$         & $~~~8.9$                             & $1.99$                      & $1.00$                           & $9.9$                                & $16.5$                                 &                              1 &                              1 &                           1 \\
 SCR~J0533-4257               & dM                                  & M4.5                                 & 05:33:28.01              & $-$42:57:18.93            & $~-17.58$                                    & $~~~~39.51$                        & $~~~96.63 \pm 0.34$         & $~~~1.4$                             & $0.85$--$6.32$              & $0.57$--$1.00$                   & $9.1$--$10.0$                        & $14.0$--$14.9$                         &                              7 &                              4 &                          12 \\
 $[$W60$]$~D43                & dM                                  & M                                    & 05:35:36.96              & $-$67:05:23.00            &                                              &                                    &                             &                                      & $1.41$                      & $0.74$                           &                                      &                                        &                              1 &                              1 &                          10 \\
 Ross~~614                    & dM                                  & M4.5V                                & 06:29:24.31              & $-$02:49:03.32            & $~~~750.14$                                  & $-802.95$                          & $~~242.97 \pm 0.88$         & $~~~0.8$                             & $2.75$--$3.84$              & $0.39$--$0.55$                   & $9.3$--$9.5$                         & $13.7$--$13.9$                         &                              2 &                              2 &                           2 \\
 Gaia~DR3~2930889294867085440 & WD                                  &                                      & 07:15:31.16              & $-$19:39:53.82            & $~~-6.06$                                    & $~~~~16.22$                        & $~~~~5.52 \pm 0.21$         & $~~~0.03$                            & $12.39$                     & $0.43$                           & $16.0$                               & $17.7$                                 &                              1 &                              1 &                           1 \\
 alf~Gem                      & HSB                                 & A1V+A2Vm                             & 07:34:35.49              & $+$31:53:14.63            &                                              &                                    & $~~~64.12 \pm 3.75$         & $~~~6.8$                             & $1.35$                      & $0.78$                           & $8.3$                                & $14.6$                                 &                              1 &                              1 &                           1 \\
 k02~Pup                      & MCP                                 & B5IV                                 & 07:38:49.74              & $-$26:48:12.84            & $~-17.54$                                    & $~~~~21.36$                        & $~~~~8.81 \pm 0.20$         & $~~~5.3$                             & $2.93$                      & $0.76$                           & $10.5$                               & $16.7$                                 &                              1 &                              1 &                           1 \\
 YZ~CMi                       & dM                                  & M4.0Ve                               & 07:44:39.66              & $+$03:33:01.74            & $-348.10$                                    & $-445.88$                          & $~~167.02 \pm 0.06$         & $~~~1.1$                             & $2.66$                      & $1.00$                           & $9.3$                                & $14.1$                                 &                              1 &                              1 &                           1 \\
 HD~~67951                    & MCP                                 & ApSiCr                               & 08:08:23.62              & $-$45:47:43.16            & $~~-4.29$                                    & $~~~~~9.96$                        & $~~~~2.65 \pm 0.04$         & $~~~1.8$                             & $3.13$                      & $0.40$                           & $12.6$                               & $17.7$                                 &                              1 &                              1 &                           1 \\
 G~~41-14                     & dM                                  & M3.5V                                & 08:58:56.75              & $+$08:28:19.76            &                                              &                                    & $~~147.66 \pm 1.98$         & $~~~0.8$                             & $19.19$                     & $0.89$                           & $10.6$                               & $15.0$                                 &                              1 &                              1 &                           1 \\
 HD~~77653                    & MCP                                 & ApSi                                 & 09:01:44.44              & $-$52:11:19.82            &                                              &                                    & $~~~~8.85 \pm 0.42$         & $~~~4.7$                             & $7.06$                      & $0.67$                           & $11.0$                               & $17.0$                                 &                              1 &                              1 &                           1 \\
 WT~2458                      & dM                                  & M4.5                                 & 09:45:57.98              & $-$32:53:26.40            & $-300.41$                                    & $~~~146.02$                        & $~~~83.23 \pm 0.09$         & $~~~0.7$                             & $1.54$                      & $0.83$                           & $10.1$                               & $14.4$                                 &                              1 &                              1 &                           1 \\
 PM~J09551-0819               & dM                                  & M1e                                  & 09:55:09.49              & $-$08:19:23.93            & $-135.78$                                    & $~~-7.16$                          & $~~~31.49 \pm 0.04$         & $~~~0.8$                             & $1.81$                      & $0.84$                           & $10.9$                               & $15.3$                                 &                              1 &                              1 &                          25 \\
 LP~~610-59                   & dM                                  &                                      & 10:43:37.55              & $-$00:48:09.06            & $-184.94$                                    & $~~~~14.43$                        & $~~~19.53 \pm 0.15$         & $~~~0.6$                             & $2.05$                      & $0.76$                           & $11.6$                               & $15.8$                                 &                              1 &                              1 &                          24 \\
 WISEA~J105315.25-085941.5    & dM                                  &                                      & 10:53:15.20              & $-$08:59:42.22            & $~~~~30.96$                                  & $~~-0.67$                          & $~~~29.38 \pm 0.06$         & $~~~1.0$                             & $1.07$--$3.43$              & $0.59$--$0.95$                   & $10.5$--$11.0$                       & $15.2$--$15.7$                         &                              3 &                              1 &                          11 \\
 ksi~UMa                      & IB                                  & F8.5:V+G2V                           & 11:18:10.18              & $+$31:31:31.12            & $-339.40$                                    & $-607.89$                          & $~~114.49 \pm 0.43$         & $~~11.3$                             & $1.49$                      & $0.68$                           & $7.4$                                & $14.1$                                 &                              1 &                              1 &                           1 \\
 WISEA~J114020.71-330519.4    & dM                                  &                                      & 11:40:20.68              & $-$33:05:19.22            & $~-62.94$                                    & $~-29.30$                          & $~~~12.94 \pm 0.08$         & $~~~0.9$                             & $2.54$                      & $0.63$                           & $11.7$                               & $16.3$                                 &                              1 &                              1 &                           1 \\
\hline
\end{tabular}

\end{threeparttable}}
  \end{table}
\end{landscape}

\begin{landscape}
  \begin{table}
    \centering \caption{Table \ref{tab:startable} continued.}
            \resizebox{\columnwidth}{!}{ 
\begin{threeparttable}
\begin{tabular}{lclrrrrccccccccc}
\hline
 \multicolumn{1}{c}{Name}   & \multicolumn{1}{c}{Variable Type}   & \multicolumn{1}{c}{Spectral Class}   & \multicolumn{1}{c}{RA}   & \multicolumn{1}{c}{Dec}   & \multicolumn{1}{c}{$\mu_\alpha\cos\delta$}   & \multicolumn{1}{c}{$\mu_\delta$}   & \multicolumn{1}{c}{$\pi$}   & \multicolumn{1}{c}{$R_e$\tnote{a}}   & \multicolumn{1}{c}{$S_I$}   & \multicolumn{1}{c}{$\fracpol$}   & \multicolumn{1}{c}{$\log_{10}T_B$}   & \multicolumn{1}{c}{$\log_{10}L_\nu$}   &   \multicolumn{1}{c}{$N_d[I]$} &   \multicolumn{1}{c}{$N_d[V]$} &   \multicolumn{1}{c}{$N_o$} \\
&&&&&\si{mas\per\year}&\si{mas\per\year}&\si{mas}&$R_\odot$&\si{\mjpb}&&\si{\kelvin}&\si{\radlum}&&&\\
\hline
 PM~J11422-0122             & dM                                  & M5e                                  & 11:42:12.84              & $-$01:22:05.61            & $~-98.06$                                    & $~~-4.86$                          & $~~~54.07 \pm 0.66$         & $~~~0.3$                             & $2.31$                      & $0.41$                           & $11.3$                               & $15.0$                                 &                              1 &                              1 &                          14 \\
 HD~105382                  & MCP                                 & B5V                                  & 12:08:05.06              & $-$50:39:40.79            & $~-34.35$                                    & $~-11.51$                          & $~~~~9.85 \pm 0.32$         & $~~~5.0$                             & $2.28$                      & $0.59$                           & $10.4$                               & $16.4$                                 &                              1 &                              1 &                           1 \\
 WISEA~J122501.45-521614.6  & dM                                  &                                      & 12:25:01.37              & $-$52:16:14.58            & $~-30.86$                                    & $~-11.29$                          & $~~~~8.14 \pm 0.30$         & $~~~0.4$                             & $3.40$                      & $0.86$                           & $13.0$                               & $16.8$                                 &                              1 &                              1 &                           1 \\
 WISEA~J123623.17-074508.2  & dM                                  &                                      & 12:36:23.03              & $-$07:45:06.81            & $~-83.14$                                    & $~-34.91$                          & $~~~25.17 \pm 0.06$         & $~~~0.7$                             & $2.28$                      & $0.60$                           & $11.2$                               & $15.6$                                 &                              1 &                              1 &                          10 \\
 UCAC4~129-071513           & YSO                                 &                                      & 12:52:22.67              & $-$64:18:38.92            & $~-38.42$                                    & $~-14.69$                          & $~~~~9.70 \pm 0.06$         & $~~~2.0$                             & $1.38$--$3.36$              & $0.84$--$1.00$                   & $11.0$--$11.4$                       & $16.2$--$16.6$                         &                              2 &                              2 &                           2 \\
 HD~115247                  & HSB                                 & F5V                                  & 13:16:02.96              & $-$05:40:07.91            & $~-83.66$                                    & $~-40.39$                          & $~~~~8.41 \pm 0.06$         & $~~10.4$                             & $2.23$--$9.87$              & $0.06$--$0.53$                   & $9.9$--$10.5$                        & $16.6$--$17.2$                         &                             20 &                              1 &                          21 \\
 BH~CVn                     & IB                                  & A6m                                  & 13:34:47.90              & $+$37:10:56.77            & $~~~~85.61$                                  & $~~-9.71$                          & $~~~21.67 \pm 0.16$         & $~~~9.8$                             & $6.64$--$7.16$              & $0.35$--$0.37$                   & $9.6$--$9.6$                         & $16.2$--$16.3$                         &                              2 &                              2 &                           2 \\
 V851~Cen                   & IB                                  & K0III                                & 13:44:00.95              & $-$61:21:58.92            & $~~~~21.07$                                  & $~~~~13.80$                        & $~~~13.50 \pm 0.04$         & $~~10.6$                             & $7.68$                      & $0.69$                           & $10.0$                               & $16.7$                                 &                              1 &                              1 &                           1 \\
 CU~Vir                     & MCP                                 & ApSi                                 & 14:12:15.72              & $+$02:24:34.21            & $~-43.05$                                    & $~-26.08$                          & $~~~13.94 \pm 0.26$         & $~~~6.3$                             & $2.17$--$13.83$             & $0.16$--$0.64$                   & $9.9$--$10.7$                        & $16.1$--$16.9$                         &                             11 &                              1 &                          11 \\
 HD~124498                  & dK                                  & K5.5Vkee                             & 14:14:21.74              & $-$15:21:17.96            & $-128.33$                                    & $-164.99$                          & $~~~34.53 \pm 0.18$         & $~~~2.3$                             & $2.17$                      & $0.62$                           & $10.0$                               & $15.3$                                 &                              1 &                              1 &                           1 \\
 WISEA~J141443.13+261639.7  & dM                                  &                                      & 14:14:43.16              & $+$26:16:40.45            & $~~~~55.60$                                  & $~~-3.50$                          & $~~~21.26 \pm 0.06$         & $~~~0.8$                             & $2.65$--$3.08$              & $0.62$--$0.72$                   & $11.3$--$11.4$                       & $15.8$--$15.9$                         &                              2 &                              2 &                           2 \\
 G~165-61                   & dM                                  & M4.5Ve                               & 14:17:02.04              & $+$31:42:44.31            &                                              &                                    & $~~~60.00 \pm 2.20$         & $~~~0.6$                             & $2.02$                      & $0.96$                           & $10.6$                               & $14.8$                                 &                              1 &                              1 &                           1 \\
 G~124-43                   & dM                                  & M4.65                                & 14:27:55.67              & $-$00:22:26.29            & $-361.50$                                    & $~~~~41.64$                        & $~~~57.08 \pm 0.06$         & $~~~1.0$                             & $1.30$--$2.97$              & $0.40$--$1.00$                   & $10.0$--$10.4$                       & $14.7$--$15.0$                         &                              9 &                              5 &                          22 \\
 LP~~325-68                 & dM                                  & M4.5Ve                               & 14:37:18.08              & $+$26:53:00.87            & $-227.64$                                    & $~~~~84.35$                        & $~~~27.29 \pm 0.06$         & $~~~0.8$                             & $3.01$                      & $0.55$                           & $11.2$                               & $15.7$                                 &                              1 &                              1 &                           1 \\
 HD~142184                  & MCP                                 & B2V                                  & 15:53:55.82              & $-$23:58:41.33            & $~-13.47$                                    & $~-23.97$                          & $~~~~7.08 \pm 0.14$         & $~~11.7$                             & $7.84$                      & $0.22$                           & $10.5$                               & $17.3$                                 &                              1 &                              1 &                           1 \\
 GSS~35                     & YSO                                 & B3                                   & 16:26:34.11              & $-$24:23:28.17            & $~~-2.17$                                    & $~-23.56$                          & $~~~~8.17 \pm 0.11$         & $~~~2.5$                             & $8.18$                      & $0.23$                           & $11.7$                               & $17.2$                                 &                              1 &                              1 &                           1 \\
 EMSR~~~20                  & YSO                                 & G7                                   & 16:28:32.50              & $-$24:22:45.81            & $~~-9.47$                                    & $~-28.06$                          & $~~~~7.35 \pm 0.06$         & $~~~2.4$                             & $2.24$                      & $0.72$                           & $11.3$                               & $16.7$                                 &                              1 &                              1 &                           1 \\
 CD-38~11343                & dM                                  & M3Ve+M4Ve                            & 16:56:48.49              & $-$39:05:38.96            & $~~~~47.00$                                  & $-113.26$                          & $~~~63.63 \pm 0.08$         & $~~~1.3$                             & $7.29$                      & $0.29$                           & $10.4$                               & $15.3$                                 &                              1 &                              1 &                           1 \\
 PM~J17021-2740             & dM                                  &                                      & 17:02:07.97              & $-$27:40:28.71            & $-129.85$                                    & $-220.40$                          & $~~~41.86 \pm 0.09$         & $~~~0.9$                             & $3.99$                      & $0.68$                           & $10.9$                               & $15.4$                                 &                              1 &                              1 &                           1 \\
 Ross~~867                  & dM                                  & M4.5V                                & 17:19:52.68              & $+$26:30:09.94            & $-226.13$                                    & $~~~355.28$                        & $~~~92.97 \pm 0.06$         & $~~~0.9$                             & $3.79$                      & $0.67$                           & $10.2$                               & $14.7$                                 &                              1 &                              1 &                           1 \\
 G~183-10                   & dM                                  & M3.5Ve                               & 17:53:00.25              & $+$16:54:59.05            & $-243.94$                                    & $-240.05$                          & $~~~44.15 \pm 1.02$         & $~~~0.8$                             & $2.09$                      & $0.51$                           & $10.6$                               & $15.1$                                 &                              1 &                              1 &                           1 \\
 2MASS~J18040683-1211342    & dM                                  &                                      & 18:04:06.89              & $-$12:11:35.71            & $~-28.06$                                    & $~~-6.91$                          & $~~~18.52 \pm 0.04$         & $~~~0.9$                             & $2.71$                      & $0.64$                           & $11.4$                               & $16.0$                                 &                              1 &                              1 &                           1 \\
 UCAC3~152-281176           & dM                                  & M5                                   & 18:45:00.95              & $-$14:09:04.59            & $~~~~47.60$                                  & $~-85.72$                          & $~~~54.97 \pm 0.09$         & $~~~0.9$                             & $5.73$                      & $0.79$                           & $10.8$                               & $15.4$                                 &                              1 &                              1 &                           2 \\
 SCR~J1928-3634             & dM                                  &                                      & 19:28:33.74              & $-$36:34:39.21            & $~~~~95.72$                                  & $-453.66$                          & $~~~39.55 \pm 0.06$         & $~~~0.7$                             & $2.51$                      & $0.64$                           & $10.9$                               & $15.3$                                 &                              1 &                              1 &                           1 \\
 2MASS~J19551247+0045365    & WD                                  &                                      & 19:55:12.45              & $+$00:45:36.33            & $~~-2.55$                                    & $~-29.55$                          & $~~~~5.85 \pm 0.08$         & $~~~0.07$                            & $2.19$                      & $0.69$                           & $14.6$                               & $16.9$                                 &                              1 &                              1 &                          13 \\
 bet~Aql~B                  & dM                                  & M3                                   & 19:55:18.83              & $+$06:24:28.14            & $~~~~22.62$                                  & $-472.30$                          & $~~~73.61 \pm 0.05$         & $~~~1.2$                             & $2.85$                      & $0.74$                           & $9.9$                                & $14.8$                                 &                              1 &                              1 &                           1 \\
 SCR~J2009-0113             & dM                                  & M5.0V                                & 20:09:18.16              & $-$01:13:45.36            & $~-43.63$                                    & $-373.73$                          & $~~~92.52 \pm 0.12$         & $~~~0.6$                             & $1.86$--$10.30$             & $0.73$--$0.92$                   & $10.3$--$11.0$                       & $14.4$--$15.2$                         &                              5 &                              5 &                          12 \\
 LEHPM~2-783                & dM                                  & M6                                   & 20:19:49.81              & $-$58:16:49.89            & $~-24.64$                                    & $-340.82$                          & $~~~61.18 \pm 0.10$         & $~~~1.0$                             & $2.43$                      & $0.81$                           & $10.2$                               & $14.9$                                 &                              1 &                              1 &                          12 \\
 2MASS~J20390476-4117390    & dM                                  &                                      & 20:39:04.97              & $-$41:17:38.95            & $~~~~58.05$                                  & $~-47.38$                          & $~~~25.63 \pm 0.06$         & $~~~0.7$                             & $1.67$                      & $0.99$                           & $11.1$                               & $15.5$                                 &                              1 &                              1 &                          12 \\
 Ross~~776                  & dM                                  & M3.3V                                & 21:16:06.14              & $+$29:51:52.37            & $~~~204.61$                                  & $~~~~24.16$                        & $~~~49.21 \pm 0.24$         & $~~~1.3$                             & $6.44$                      & $0.50$                           & $10.6$                               & $15.5$                                 &                              1 &                              1 &                           1 \\
 UCAC3~89-412162            & dM                                  &                                      & 21:22:17.56              & $-$45:46:31.42            & $~-53.47$                                    & $~-34.00$                          & $~~~31.55 \pm 0.04$         & $~~~0.8$                             & $2.31$                      & $0.87$                           & $11.0$                               & $15.4$                                 &                              1 &                              1 &                          12 \\
 HD~210547A                 & HSB                                 &                                      & 22:13:00.08              & $-$61:58:53.05            & $~-13.18$                                    & $~~~~~0.79$                        & $~~~~5.82 \pm 0.03$         & $~~~5.7$                             & $1.44$                      & $0.94$                           & $10.5$                               & $16.7$                                 &                              1 &                              1 &                           9 \\
 Wolf~1561~A                & dM                                  & M4V                                  & 22:17:18.29              & $-$08:48:18.13            & $-458.51$                                    & $-291.85$                          & $~~~89.80 \pm 0.09$         & $~~~0.7$                             & $1.74$--$1.85$              & $0.69$--$1.00$                   & $10.0$--$10.1$                       & $14.4$--$14.4$                         &                              2 &                              1 &                          14 \\
 LAMOST~J221747.87-030039.0 & dM                                  & M5                                   & 22:17:47.86              & $-$03:00:38.82            & $~~~~21.79$                                  & $~~-9.96$                          & $~~~~6.71 \pm 0.08$         & $~~~1.3$                             & $5.50$                      & $0.77$                           & $12.2$                               & $17.2$                                 &                              1 &                              1 &                          36 \\
 UPM~J2224-5826             & dM                                  &                                      & 22:24:24.51              & $-$58:26:14.13            & $~-75.71$                                    & $~-37.18$                          & $~~~62.38 \pm 0.19$         & $~~~1.5$                             & $4.12$--$5.45$              & $0.85$--$1.00$                   & $10.1$--$10.2$                       & $15.1$--$15.2$                         &                              2 &                              1 &                          20 \\
 SCR~J2241-6119A            & dM                                  &                                      & 22:41:44.76              & $-$61:19:33.18            & $~~~150.23$                                  & $~-87.90$                          & $~~~35.19 \pm 0.05$         & $~~~0.9$                             & $1.38$--$5.44$              & $0.39$--$1.00$                   & $10.6$--$11.2$                       & $15.1$--$15.7$                         &                              6 &                              4 &                           9 \\
 SZ~Psc                     & IB                                  & G5Vp                                 & 23:13:23.70              & $+$02:40:33.01            & $~~~~27.03$                                  & $~~~~27.33$                        & $~~~11.10 \pm 0.08$         & $~~13.7$                             & $2.07$--$11.89$             & $0.03$--$0.84$                   & $9.4$--$10.1$                        & $16.3$--$17.1$                         &                             16 &                              1 &                          16 \\
\hline
\end{tabular}
\begin{tablenotes} \item[a] $R_e$ taken as: $3R_\star$ for
            single stars as calculated from Gaia DR3 photometry \citep{Gaia2022} and the
            inter-binary region for interacting binaries.  \end{tablenotes}
\end{threeparttable}}
  \end{table}
\end{landscape}

\begin{figure}
  \centering
  \includegraphics[width=0.5\textwidth]{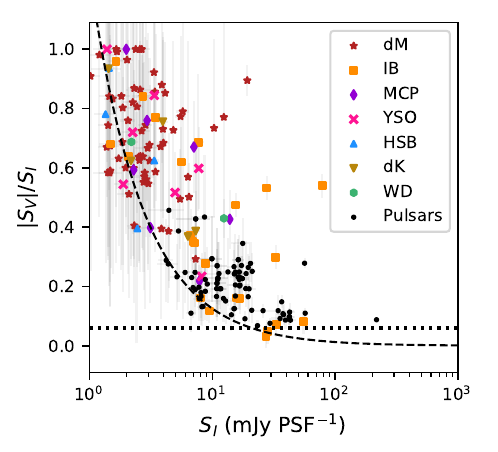}
  \caption{
    \small
    Fractional circular polarisation and flux density phase space, with the
    radio stars in our sample labelled according to variable class. M-dwarfs (dM) are shown as
    red stars, K-dwarfs (dK) as brown stars, interacting binaries (IB) as yellow squares,
    magnetic chemically peculiar (MCP) stars as purple diamonds, young stellar objects (YSO) as
    fuchsia crosses, hot spectroscopic binaries (HSB) as blue triangles, and white dwarfs (WD)
    as green hexagons. The dotted line indicates a fractional circular polarisation of
    \SI{6}{\percent} below which we excluded candidates from initial classification, retaining
    only the low $\fracpol$ detections of previously classified sources. The dashed line
    indicates a $5\sigma_V$ detection threshold where $\sigma_V = \SI{0.25}{\mjpb}$ is the
    median RMS noise in Stokes $V$, and represents the average sensitivity limit in fractional
    polarisation across our survey. Local variations in RMS result in a few detections below
    this threshold.
  }\label{fig:fp-flux-phasespace}
\end{figure}

In \cref{fig:fp-flux-phasespace} we show the fractional circular polarisation $\fracpol$ as a
function of Stokes~$I$ flux density for all star detections in RACS-low and VASTP-low. We also
show detected pulsars that are well separated in this space from the majority of single star
detections, having lower $\fracpol$ of ${\sim}\SIrange{10}{30}{\percent}$. This is slightly
higher than the fractional circular polarisation of pulsars detected in LoTSS by
\citet{Callingham2023}, who find most pulsars below the $\fracpol \sim \SI{10}{\percent}$
level, however this is likely just a selection effect due to the
$\fracpol \sim \SI{6}{\percent}$ limit of our search. For a more complete analysis of known
pulsars detected in RACS-low we refer the reader to \citet{Anumarlapudi2023}, who find a mean
pulsar $\fracpol$ of $\sim \SI{5}{\percent}$.

\begin{figure}
  \centering
  \includegraphics[width=0.5\textwidth]{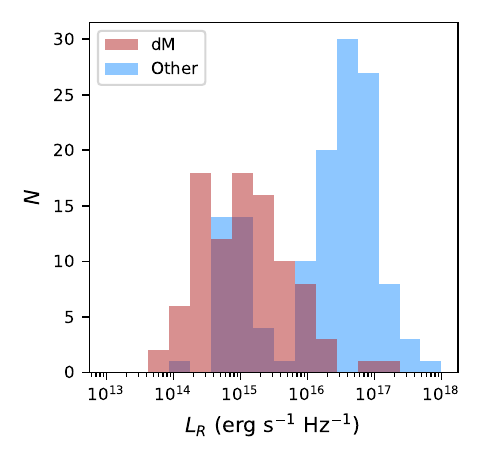}
  \caption{
    \small
    Stokes $I$ radio luminosity distribution of all star detections in RACS-low
    and VASTP-low. M-dwarfs are shown in red and all other stars are combined and shown in
    blue, as the sample sizes of individual variable and spectral classes are too low to
    meaningfully show them separately.
  }\label{fig:lrad-hist}
\end{figure}

\begin{figure}
  \centering
  \includegraphics[width=0.5\textwidth]{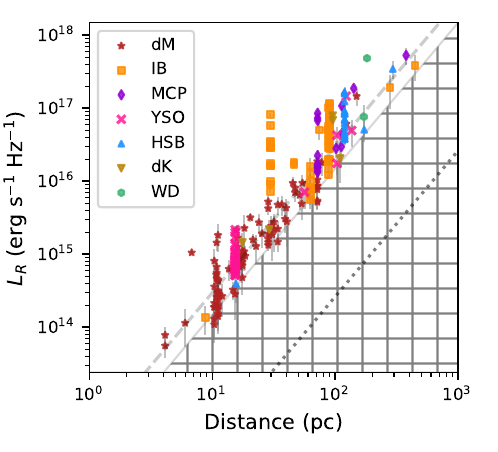}
  \caption{
    \small
    Stokes $I$ radio luminosity of each star detection as a function of
    distance. The hatched region corresponds to Stokes $I$ flux densities below the $5\sigma$
    detection threshold of \SI{1.25}{\mjpb}. The dashed line indicates the $5\sigma$ Stokes $V$
    detection threshold for $\fracpol = 0.5$, with lower $\fracpol$ shifting the threshold to
    higher $L_\nu$. The dotted line represents the projected $5\sigma$ detection threshold of
    \SI{22}{\micro\jansky\per\beam} in a \SI{1}{\hour} observation with SKA-mid
    \citep{Braun2019}. Repeat detections are visible as vertically aligned points.
  }\label{fig:lrad-dist}
\end{figure}

In \cref{fig:lrad-hist} we show the Stokes $I$ radio luminosity distribution of our sample,
with M-dwarfs tending to group around $L_\nu = \SI{e15}{\radlum}$ while other variable types
group around $L_\nu = \SIrange{e16}{e17}{\radlum}$. Radio luminosity is also shown as a
function of stellar distance in \cref{fig:lrad-dist}. The hatched region represents flux
densities below our $5\sigma$ detection threshold of \SI{1.25}{\mjpb} in Stokes $I$, which
preferentially suppresses counts at the low end of the $L_\nu$ distribution. The completeness
of our sample is further dependent upon fractional circular polarisation, with lower $\fracpol$
shifting the detection threshold to higher $L_\nu$, indicated by the dashed line at
$\fracpol = 0.5$.

To place constraints on plausible radio emission mechanisms we calculated the brightness
temperature

\begin{equation}
  T_B = \frac{Sc^2}{2\pi k_B \nu^2}\frac{d^2}{R_e^2}
\end{equation}

where $k_B$ is the Boltzmann constant, $\nu$ is the observing frequency of
\SI{887.5}{\mega\hertz}, $d$ is the stellar distance, and $R_e$ is the length scale of the
emission region. We have assumed an upper limit of $R_e = 3R_\star$ for each star, and
additionally calculate $T_B$ for interacting binaries with $R_e$ equal to three times the
binary separation distance. The true emission region is likely much smaller than these limits,
particularly for detections with large $\fracpol$ as they require a consistent orientation of
the magnetic field within the emission region to produce predominantly right or left handed
circular polarisation.

\begin{figure*}
  \centering
  \includegraphics[width=\textwidth]{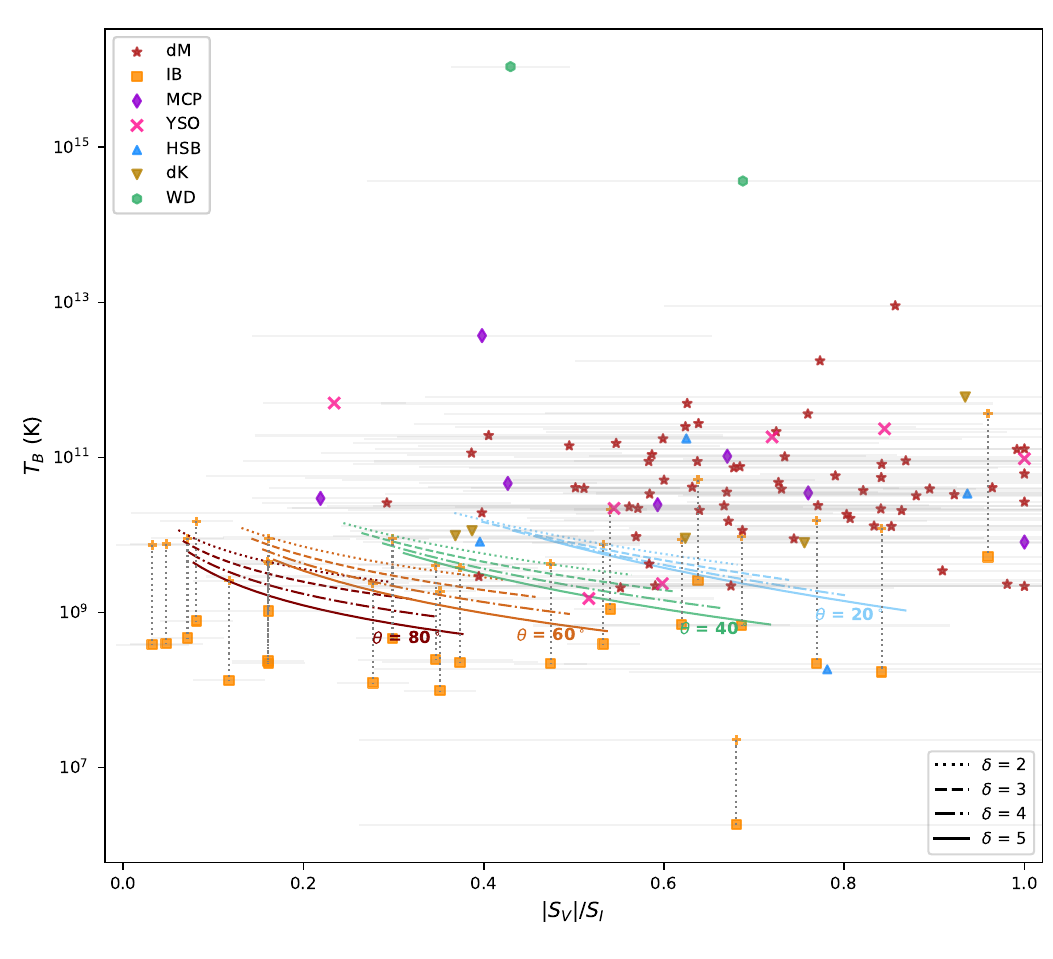}
  \caption{ \small Brightness temperature and fractional circular polarisation phase space,
    with stars labelled according to variable class (see
    \cref{fig:fp-flux-phasespace}). Vertical dotted lines connect two $T_B$ estimates for
    interacting binary systems, with estimated emission regions of the inter-binary region and
    stellar disk represented as yellow squares and crosses respectively. Empirical models of
    the maximum effective temperature of optically thin gyro-synchrotron emission
    \citep{Dulk1985} are shown for a range of electron energy index $\delta$ and viewing angle
    $\theta$.  }\label{fig:tb-fp-phasespace}
\end{figure*}

In \cref{fig:tb-fp-phasespace} we show all star detections in a brightness
temperature--fractional circular polarisation phase space alongside empirical models of the
maximum effective temperature of optically thin gyrosynchrotron emission \citep{Dulk1985}. Most
interacting binary systems feature minimum $T_B$ of \SIrange{e6}{e10}{\kelvin} and
$\fracpol < 0.8$ and are plausibly driven by gyrosynchrotron emission which is commonly
detected from RS CVn systems, though we note that coherent emission from the ECMI is also
possible \citeg{Slee2008, Toet2021}. The majority of other detections including one RS CVn,
however, feature minimum $T_B$ of order \SIrange{e10}{e16}{\kelvin}, and as these detections
feature high fractional circular polarisation the emission necessarily originates from a
compact region in which the magnetic field is nearly uniform. This constraint on the size of
the emission region further increases the minimum brightness temperature, and combined with the
large fractional circular polarisation makes gyrosynchrotron emission unlikely. These
detections are therefore likely driven by a coherent emission process such as the ECMI or
plasma emission.

\section{POPULATION STATISTICS}\label{sec:popstats}

\subsection{Detection Counts}\label{sec:detection-counts}

We detected a total of 36 stars within the VASTP-low footprint over 15 epochs. Due to slight
variations in the survey footprint between epochs and overlap between neighbouring tiles, each
star was observed between 9--36 times with a median of 14 observations. To characterise the
repeat detection rates of each star we calculate the detection fraction $DF = N_d/N_o$ where
$N_o$ and $N_d$ are the respective number of observations and detections of each star. These
counts only consider observations in which the local 3-sigma RMS noise is less than the flux
density of the weakest detection, removing observations in which the star is located toward the
image edges or corners where sensitivity degrades.

\begin{figure}
  \centering
  \includegraphics[width=0.5\textwidth]{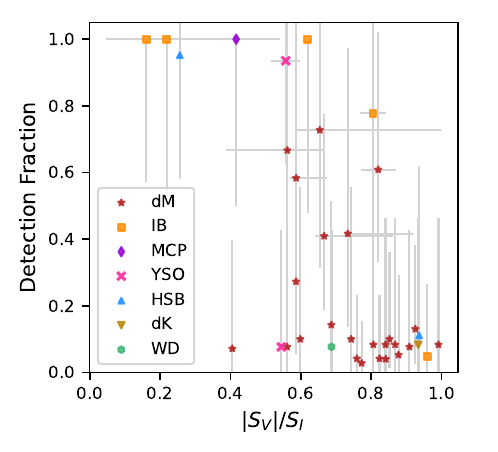}
  \caption{
    \small
    Detection fraction as a function of fractional circular polarisation for our
    sample of repeat observed stars in the VASTP-low footprint. The $\fracpol$ error bars
    indicate the range of observed fractional circular polarisation for each star, and the
    detection fraction error bars are derived from \SI{95}{\percent} Poisson confidence
    intervals on the number of detections. Markers denote the sub-type of variable star
    (see \cref{fig:fp-flux-phasespace}).
  }\label{fig:fp-dutycycle-phasespace}
\end{figure}

In \cref{fig:fp-dutycycle-phasespace} we show the detection fraction as a function of
fractional polarisation. Most of the stars with high $DF$ are interacting binaries that feature
a quiescent component to the radio emission, along with one YSO (AB Dor) and one MCP star (CU
Vir) that are both well-known, persistent radio emitters. We find the majority of highly
circularly polarised detections are associated with single detections of M-dwarfs with a median
$DF$ of \dfmean, and take this as an upper limit on the average population detection fraction
$\mean{DF}$. This upper limit is consistent with previous estimates of the duty cycle of both
stochastic radio bursts \citep{Villadsen2019} and rotationally modulated auroral emission
\citep{Hallinan2007, Nichols2012}.

\subsection{Radio Activity Fraction of the M-Dwarf Population}\label{sec:activity-fraction}

As we detect new stars in each successive analysed epoch, an upper limit to the detection
fraction implies the existence of low burst-rate stars within the population that produce
detectable radio bursts but have not been sufficiently sampled to produce a detection. We
define a burst to be detectable if it has a radio luminosity sufficient to be detected above
our $5\sigma$ detection threshold of \SI{1.25}{\mjpb}, which corresponds to a detectability
horizon $d_{\rm max}$ for a particular burst radio luminosity. In this section we use the
repeat detection rates of stars within the VASTP-low footprint to estimate the size of the
detectable population. We limit our analysis to M-dwarfs due to the low sample size of other
variable and spectral classes. We further restrict our sample to $d_{\rm max} < \horizon$
within which bursts above the typical M-dwarf radio luminosity of \SI{e15}{\radlum} are
detectable, as our sample becomes increasingly incomplete in radio luminosity at greater
distances.

We assume there are $n$ detectable M-dwarfs per unit solid angle, each with a burst rate
$\lambda$ sampled from the population distribution $f(\lambda)$ such that
$nAf(\lambda)\diff{\lambda}$ is the number of detectable M-dwarfs with burst rates between
$\lambda$ and $\lambda + \diff{\lambda}$ in a solid angle $A$. For a star with rate $\lambda$
the number of detectable bursts in an epoch of duration $\Delta t$ is $\lambda \Delta t$. Hence
the expected number of bursts detected in the epoch from all stars in a solid angle $A$ is

\begin{align}
  N_d &= \int_0^\infty nA \lambda \Delta t f(\lambda) \diff{\lambda} \\
      &= nA \Delta t \mean{\lambda}
\end{align}

where $\mean{\lambda} = \int_0^\infty \lambda f(\lambda) \diff{\lambda}$ is the average burst
rate over the observed stars. The expected number of bursts in $M$ epochs is then

\begin{align}\label{eq:Nbursts}
  \sum_{i=1}^M N_{d,i} &= n \mean{\lambda} \sum_{i=1}^M A_i \Delta t_i
\end{align}

where $A_i$ and $\Delta t_i$ are the solid angle coverage and observation duration of epoch $i$.

None of our detections show evidence of bursts on timescales shorter than the observation time
so we assume each detection represents a single burst, and as our RACS-low detections are all
bright enough to be detected in a \SI{12}{\minute} observation we adopt a common
$\Delta t = \SI{12}{\minute}$ for all epochs. The average detection fraction $\mean{DF}$ is
then an estimate of the average burst rate:

\begin{equation}\label{eq:mean-lambda-df}
  \mean{\lambda} = \frac{\mean{DF}}{\Delta t}, 
\end{equation}

and the total number of detectable M-dwarfs in the VASTP-low footprint of area
$A_S = \surveyarea$ is then, using \cref{eq:Nbursts,eq:mean-lambda-df}:

\begin{align}\label{eq:Nstars}
  \mathcal{N} = nA_S &=  \frac{A_S}{{\sum_{i=1}^M A_i}} \frac{1}{\mean{DF}} \sum_{i=1}^M N_{d,i}.
\end{align}

Within \horizon\ we detect 11 M-dwarfs a total of 46 times. Using the solid angle coverage of
each epoch listed in \cref{tab:epochs} with \cref{eq:Nstars} we estimate at least
$\mathcal{N}=$ \dkmstarprediction\ M-dwarfs within the VASTP-low footprint should produce
detectable radio bursts, where the uncertainties are derived from a \SI{95}{\percent} Poisson
confidence interval on the detection counts. We compare this to the total number of M-dwarfs in
this volume from the Fifth Catalogue of Nearby Stars \citep[CNS5;][]{Golovin2023}. This
catalogue is statistically complete out to \horizon\ for spectral types earlier than L8 and
contains 471 M-dwarfs within the VASTP-low footprint. Our analysis therefore indicates that at
least \dkmactivityfraction\ of M-dwarfs should produce radio bursts more luminous than
\SI{e13}{\radlum}. Given the footprint area of \surveyarea\ this estimate corresponds to a
surface density of $n=$ \longtermdkmdensity, such that at least one M-dwarf within \horizon\
should be detectable per \SI{66}{\deg^2} low-band field.

These results are derived for our sample within \horizon, while including the more distant high
luminosity bursts we detect twice as many events. Due to incompleteness in the luminosity
distribution below \SI{e15}{\radlum} and the $DF$ distribution below $DF \sim 0.07$, these
values should be interpreted as lower limits. Extending this analysis to future epochs in the
full VAST survey will push the $DF$ incompleteness to lower thresholds and allow sampling of
stars with lower burst rates, probing the true turnover in the burst rate distribution and
providing a more accurate estimate of the surface density and activity fraction of radio loud
M-dwarfs.

\subsection{Instantaneous Surface Density}\label{sec:surface-density}

\begin{figure*}
  \centering
  \includegraphics[width=\textwidth]{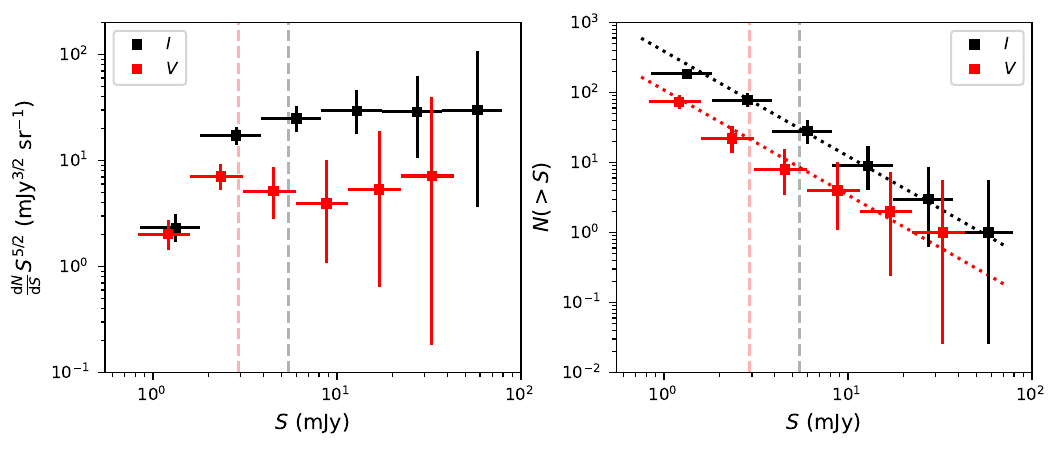}
  \caption{
    \small
    Differential (left) and cumulative (right) source counts for all star detections in
    RACS-low and VASTP-low, with Stokes~$I$ and $V$ shown in black and red respectively. The
    red dashed line represents the RACS-low \SI{95}{\percent} completeness limit
    \citep{Hale2021} which we take as an estimate of the Stokes~$V$ completeness limit in our
    survey, and the black dashed line is the corresponding Stokes~$I$ completeness limit for a
    median fractional polarisation of \SI{50}{\percent}. Differential counts are multiplied by
    a $S^{5/2}$ Euclidean normalisation, and the cumulative counts are shown alongside dotted
    lines representing a Euclidean source distribution with power law slope of
    $N(>S) \propto S^{-3/2}$.
  }\label{fig:logN-logS}
\end{figure*}

In \cref{fig:logN-logS} we show the differential and cumulative source counts for all star
detections in our sample in Stokes~$I$ and $V$, with detections combined into six
logarithmically spaced flux density bins between \SIrange{1}{60}{\milli\jansky}. Counts in both
polarisations are reduced at low flux densities due to incompleteness. RACS-low sources in
Stokes~$I$ have a \SI{95}{\percent} completeness limit of \SI{2.9}{\milli\jansky}
\citep{Hale2021} which we take as an estimate of the \SI{95}{\percent} Stokes~$V$ completeness
limit, indicated by the dashed red line.\footnote{The Stokes~$V$ RMS noise is slightly lower
  than in Stokes~$I$, particularly towards poorly deconvolved bright sources or highly confused
  regions with diffuse Galactic emission, and the true Stokes~$V$ completeness limit is
  therefore lower than this estimate.} The completeness limit in Stokes~$I$ varies due to the
range of fractional polarisation in our sample, so we show the limit for the median $\fracpol$
of \SI{50}{\percent} as the black dashed line. Above these thresholds the source counts are in
agreement with a Euclidean source distribution, with flat slope in the normalised differential
counts and cumulative counts $N(>S) \propto S^{-3/2}$. This is consistent with expectations as
our sample is contained within $\sim$\SI{500}{\parsec} and therefore any directional bias in
counts due to Galactic plane structure is minimal \citep{West2008}.

Between RACS-low and VASTP-low our circular polarisation search has covered a total area of
\SI{135\,498}{\deg^2} and resulted in 229 radio star detections, 96 of which are of M-dwarfs
with the majority of detections attributable to coherent radio bursts. This corresponds to an
instantaneous radio star detection surface density of \stardensity\ on \SI{12}{\minute}
timescales, and \dkmdensity\ considering only M-dwarf detections. The full VAST survey with
ASKAP will run with similar observing parameters to VASTP-low with a plan to conduct 5953
\SI{12}{\minute} observations annually, and should therefore detect \vaststarcount\ radio stars
including \vastdkmstarcount\ M-dwarfs each year. Scaling our measured surface densities to the
expected SKA-mid \SI{770}{\mega\hertz} 1-hour sensitivity of \SI{4.4}{\micro\jansky\per\beam}
\citep{Braun2019} and accounting for the factor of five increase in integration time, we expect
a radio and M-dwarf surface density of \skastardensity\ and \skadkmdensity\ respectively.
All-sky surveys with SKA-mid should therefore expect to detect \skastarcount\ radio stars, with
a \SI{e15}{\radlum} completeness horizon of $\sim$\SI{200}{\parsec} and \SI{e17}{\radlum}
bursts detectable out to $\sim$\SI{2}{\kilo\parsec}.

\section{CONCLUSIONS AND OUTLOOK}

As part of the ASKAP pilot survey program we have conducted a multi-epoch circular polarisation
search for radio stars. Between RACS-low and VASTP-low we have made 229 detections of a sample
of 76 stars, with the majority of detections attributed to coherent M-dwarf radio
bursts. Through repeat observations of the VASTP-low survey footprint we constrain the typical
duty cycle of radio bursts in the M-dwarf population to less than \SI{8}{\percent}, and find
that at least \dkmactivityfraction\ of the population should produce bursts with a radio
luminosity greater than \SI{e13}{\radlum}.

Circular polarisation searches have been well demonstrated as a discovery tool for radio stars,
and full ASKAP surveys will produce hundreds of new radio star detections per year helping to
build larger statistical samples. The full VAST survey is underway, and with further repeat
sampling we will be able to better characterise the rate and luminosity distributions of
M-dwarf radio bursts and extend this analysis to other radio star classes. Repeat detections
will provide insight into the range of burst parameters in individual stars and their
relationship with other stellar activity parameters, such as rapid rotation, chromospheric
activity, and multi-wavelength transient behaviour.

These surveys will also provide a large number of candidates for targeted followup with
wide-bandwidth, high instantaneous sensitivy instruments such as the Australia Telescope
Compact Array (ATCA) and MeerKAT, allowing the phenomenology of their activity to be further
characterised with dynamic spectroscopy and polarimetry.  Identification of the emission
processes responsible for coherent stellar radio bursts will provide important insights into
the magnetospheric processes present in these systems, helping to distinguish whether they more
resemble the stochastic radio activity observed from the Sun, rotationally modulated pulses
driven by auroral current systems, or other magnetospheric processes unobserved in the Solar
system.

\section*{Acknowledgements}

JP is supported by Australian Government Research Training Program Scholarships. DK and AO are
supported by NSF grant AST-1816492. This scientific work uses data obtained from Inyarrimanha
Ilgari Bundara / the Murchison Radio-astronomy Observatory. We acknowledge the Wajarri Yamaji
People as the Traditional Owners and native title holders of the Observatory site. CSIRO’s
ASKAP radio telescope is part of the Australia Telescope National Facility
(https://ror.org/05qajvd42). Operation of ASKAP is funded by the Australian Government with
support from the National Collaborative Research Infrastructure Strategy. ASKAP uses the
resources of the Pawsey Supercomputing Research Centre. Establishment of ASKAP, Inyarrimanha
Ilgari Bundara, the CSIRO Murchison Radio-astronomy Observatory and the Pawsey Supercomputing
Research Centre are initiatives of the Australian Government, with support from the Government
of Western Australia and the Science and Industry Endowment Fund. This research made use of the
following {\sc python} packages: {\sc Astropy} \citep{Astropy2013, Astropy2018}, a
community-developed core Python package for Astronomy, {\sc matplotlib} \citep{Hunter2007}, a
Python library for publication quality graphics, {\sc NumPy} \citep{vanderWalt2011,
  Harris2020}, and {\sc pandas} \citep{McKinney2010, McKinney2011}.

\section*{Data Availability}

The ASKAP data analysed in this paper (RACS-low and VASTP-low) can be accessed through the
CSIRO ASKAP Science Data Archive
(CASDA\footnote{\url{https://data.csiro.au/domain/casdaObservation}}) under project
codes AS110 and AS107.  


\bibliographystyle{mnras}
\bibliography{bibfile}

\bsp	
\label{lastpage}
\end{document}